\documentclass{aastex}          
\usepackage{spr-astr-addons}    
\usepackage{subeqnarray}
\begin{document}
%
%
%
%
\def\astrobj#1{#1}
\newenvironment{lefteqnarray}{\arraycolsep=0pt\begin{eqnarray}}
{\end{eqnarray}\protect\aftergroup\ignorespaces}
\newenvironment{lefteqnarray*}{\arraycolsep=0pt\begin{eqnarray*}}
{\end{eqnarray*}\protect\aftergroup\ignorespaces}
\newenvironment{leftsubeqnarray}{\arraycolsep=0pt\begin{subeqnarray}}
{\end{subeqnarray}\protect\aftergroup\ignorespaces}
\newcommand{\diff}{{\rm\,d}}
\newcommand{\appleq}{\stackrel{<}{\sim}}
\newcommand{\appgeq}{\stackrel{>}{\sim}}
\newcommand{\Int}{\mathop{\rm Int}\nolimits}
\newcommand{\Nint}{\mathop{\rm Nint}\nolimits}
\newcommand{\range}{{\rm -}}
\newcommand{\displayfrac}[2]{\frac{\displaystyle #1}{\displaystyle #2}}
%
%
\title{Oxygen abundance in local disk and bulge: \\
chemical evolution with a strictly universal IMF}
%
%
\shorttitle{Oxygen abundance and chemical evolution}
\shortauthors{Caimmi and Milanese}

\author{R. Caimmi, E. Milanese
\altaffilmark{Astronomy Department, Padua University, Vicolo Osservatorio 2,\\
35122 PADOVA (Italy)}} 
\email{roberto.caimmi@unipd.it} 
\email{elena.milanese@studenti.unipd.it}


\begin{abstract}
\noindent\noindent
This paper has two parts: one about observational constraints
related to the empirical differential oxygen abundance
distribution (EDOD), and the other about inhomogeneous
models of chemical evolution, in particular the theoretical
differential oxygen abundance distribution (TDOD).
In the first part,
the EDOD is deduced from subsamples related to two different
samples involving (i) $N=532$ solar neighbourhood (SN)
stars within the range, $-1.5<$[Fe/H]$<0.5$, for
which the oxygen abundance has been determined
both in presence and in absence of the local thermodynamical
equilibrium (LTE) approximation 
\textsf{[Ramirez et al., \aap ~{\textbf 465}, 271 (2007)]};
and (ii) $N=64$ SN thick disk, SN thin disk, and
bulge K-giant
stars within the range, $-1.7<$[Fe/H]$<0.5$, for
which the oxygen abundance has been determined
\textsf{[Melendez et al., \aap ~{\textbf 484}, L21 (2008)]}.
A comparison is made with previous results implying use
of [O/H]-[Fe/H] empirical relations 
\textsf{[Caimmi, Astron. Nachr. {\textbf 322}, 241 (2001b);
Caimmi, \na ~{\textbf 12}, 289 (2007)]} 
related to (iii) 372 SN halo subdwarfs
\textsf{[Ryan and Norris, \aj ~{\bf 101}, 1865 (1991)]};
and (iv) 268 K-giant bulge stars
\textsf{[Sadler et al., \aj ~{\textbf 112}, 171 (1996)]}.
The EDOD of the SN
thick + thin disk is determined by weighting 
the mass, for assumed SN thick to thin disk mass ratio
within the range, 0.1-0.9.

In the second part, inhomogeneous models of chemical evolution
for the SN thick disk, the SN thin disk, the SN thick + thin disk,
the SN halo, and the bulge, are computed assuming the instantaneous
recycling approximation.
The EDOD data are fitted, to an acceptable extent, by their
TDOD counterparts with the exception of the thin or
thick + thin disk, where two additional restrictions
are needed:
(i) still undetected, low-oxygen abundance
thin disk stars exist, and (ii) a single
oxygen overabundant star is removed from
a thin disk subsample.
In any case, the (assumed power-law) stellar initial
mass function (IMF) is universal but
gas can be inhibited from, or enhanced in, forming stars at
different rates with respect to a selected reference case.
Models involving a strictly universal IMF (i.e. gas neither
inhibited from, nor enhanced in, forming stars with respect
to a selected reference case) can also reproduce the data
to an acceptable extent.

Our main conclusions are (1) different models are necessary
to fit the (incomplete) halo sample, which is consistent
with the idea of two distinct halo components: an inner,
flattened halo in slow prograde rotation, and an outer,
spherical halo in net retrograde rotation
\textsf{[Carollo et al., \nat ~{\textbf 450}, 1020 (2007)]};
(2) the oxygen enrichment
within the inner SN halo, the SN thick disk, and the bulge,
was similar and coeval within the same metallicity range,
as inferred from observations
\textsf{[Prochaska et al., \aj ~{\textbf 120}, 2513 (2000)]};
(3) the fit to thin disk data implies an oxygen abundance
range similar to its thick disk counterpart, with the
extension of conclusion (2) to the thin disk, and the evolution
of the thick + thin disk as a whole
\textsf{[Haywood, \mnras ~{\textbf 388}, 1175 (2008)]}
cannot be excluded; (4) leaving outside the outer halo,
a fit to the data related to different environments is
provided by models with a strictly universal IMF but
different probabilities of a region being active, which
implies different global efficiencies of the star formation
rate; (5) a special case of stellar
migration across the disk can be described by models
with enhanced star formation, where a fraction of
currently observed SN stars were born {\it in situ}
and a comparable fraction is due to the net effect
of stellar migration, according to recent results
based on high-resolution $N$-body + smooth particle
hydrodynamics simulations
\textsf{[Ro\v{s}kar et al., \apjl ~{\textbf 684}, L79 (2008)]}.

\noindent

\end{abstract}

\keywords{galaxies: evolution - stars: formation; evolution.}

\section{Introduction} \label{s:intro}

The synthesis of $\alpha$-elements (such as O, Ne,
Mg, Si, S, Ca) takes place within SnII progenitors,
via reactions of the type, E + $\alpha$ $\to$ E$^\prime$ +
$\gamma$, where E, E$^\prime$, are generic elements (E lighter
than E$^\prime$), $\alpha$ is a helium nucleus, and $\gamma$ a
photon.   Accordingly, $\alpha$-elements are primary
elements, for which the yield is not sensitive to the
initial abundance of carbon or other heavy elements
in the progenitor, because they are synthesized
directly from hydrogen and helium.    The abundances
of $\alpha$-elements as a function of the metal
content provide crucial information about the 
stellar initial mass function (IMF) and
formation history
(e.g., \cite{tin80}; \cite{pag89}; \cite{pra00};
\cite{raa07}; \cite{mea08}),
and make a main constraint on models of chemical
evolution.   Among $\alpha$-elements, oxygen has a
special role, in that it is the most abundant
element in the universe, after hydrogen and helium.

Unfortunately, oxygen is more difficult than
iron (which is not an $\alpha$-element) to detect,
and empirical relations were used
to express the former as a function of the latter
(e.g., \cite{bar88}; \cite{pag89}; \cite{abr89};
\cite{caa00}; \cite{gra00};
\cite{taa01}; \cite{isa01};
for a review refer to \cite{baa01}).   Only
recently, direct oxygen abundance determinations
on solar neighbourhood (SN) thick disk, thin
disk, and halo stars%
\footnote{In general, SN stars are to be intended
as ``passing through the solar neighbourhood''
(e.g., \cite{ryn91}).}, %
and bulge stars have been performed (e.g., \cite{raa07};
\cite{mea08}).   Different oxygen
abundance at any given iron abundance range
in common between two selected environments
implies related stars were formed from
different mixtures of gas, presumably at
different epochs and/or locations.   Oxygen
abundance patterns and their systematic
diversities between environment stars
thus constrain important information about
the assembling and evolution of the Galaxy.

In particular, the chemical abundance of
thick disk stars suggests a similar history
to those of metal-rich ([Fe/H]$\approx-1.3$)
halo stars, and the thick disk abundance
patterns show excellent agreement with the
chemical abundances observed in metal-poor
([Fe/H]$\appleq-0.4$) bulge stars, suggesting
the two populations were formed from the
same reservoir at a common epoch, regardless
of the exact physical process involved in
their formation (\cite{pra00}).
Additional support to the last conclusion
is provided by recent findings (\cite{mea08}).
On the other hand, a
different dynamical signature is exhibited
by the Galactic spheroid and disk, which
makes evidence for a distinct halo-bulge
and thick disk - thin disk collapse, as shown
by different empirical distributions of
specific angular momentum (\cite{wyg92};
\cite{ibg95}).

Concerning metal abundance distributions
in SN disk stars, it has been recognized
that earlier results based on spectral
type selection, were biased against stars
with solar metallicity or higher, with
respect to recent results based on colour
selection (Haywood (2001, 2006).
Oxygen
abundance distributions in SN disk stars,
or thick disk and thin disk separately,
can be deduced from recent direct determinations
(\cite{raa07}).   The same cannot
be done for halo and bulge stars, with the
exception of restricted samples (e.g.,
\cite{raa07}; \cite{mea08}),
and some empirical [O/H]-[Fe/H] relation
must be used (e.g., \cite{cai07}, hereafter
quoted as C07).   Recent analysis on
high-resolution infrared spectra of giant
stars, show no chemical distinction between
the bulge and the SN thick disk (\cite{mea08}).

Inhomogeneous (i.e. implying inhomogeneous
star formation) models of chemical evolution
succeed in both providing a solution to the
G-dwarf problem and reproducing substantial
scatter exhibited by the empirical SN
age-metallicity relation (e.g., \cite{maa93};
\cite{cai00}, \cite{cai01a}, hereafter
quoted together as C00; \cite{cai01b},
hereafter quoted as C01; C07; \cite{cai08},
hereafter quoted as C08).   The current
paper aims to investigate if inhomogeneous
simple models of chemical evolution with
strictly universal IMF (implying gas is neither
inhibited from, nor enhanced in, forming stars
with regard to different environments) provide
an acceptable fit to the oxygen abundance
distribution in the SN thick disk, SN thin
disk, SN halo, and bulge, for which new data
are available, even if biased towards low
metallicities or incomplete.
Though the existence of a universal IMF has
recently been questioned (e.g., \cite{bka07}),
it appears as a natural consequence of
similar (but not necessarily shared) chemical
evolution histories experienced by the
metal-rich SN halo and the SN thick disk on
one hand (\cite{pra00}), and the
bulge and the SN thick disk on the other
hand (\cite{mea08}).

The oxygen abundance is deduced from a sample
$(N=523)$ of SN disk and SN halo stars
(\cite{raa07}), hereafter quoted as
the Ra07 sample, from which four subsamples
are extracted following the prescriptions of
the authors, related to thick disk $(N=
133)$, thin disk $(N=310)$, halo $(N=28)$,
and uncertain population $(N=51)$ stars, hereafter
quoted as the RaK07, RaN07, RaH07, and RaU07
subsample, respectively.   The RaK07 subsample
is biased against low ([Fe/H]$\appleq-1$)
metallicities, and the RaH07 subsample is
incomplete (\cite{raa07}).   For the
bulge and the halo earlier results (C07) are
also used, where oxygen abundance is deduced
from iron abundance using empirical relations,
with regard to a sample $(N=268)$ of bulge K
giants in Baade$^\prime$s window (\cite{saa96}),
hereafter quoted as the Sa96 sample, and a
sample $(N=372)$ of kinematically selected
halo subdwarfs (\cite{ryn91}),
hereafter quoted as the RN91 sample.

The oxygen abundance is also deduced from a
more restricted sample $(N=68)$ of SN disk,
SN halo, and bulge stars (\cite{mea08}),
hereafter quoted as the Ma08 sample,
which is subdivided into four subsamples,
related to the thick disk $(N=21)$, thin disk
$(N=24)$, halo $(N=4)$, and bulge $(N=19)$,
hereafter quoted as the MaK08, MaN08, MaH08,
and MaB08 subsample, respectively.   All
objects have similar stellar parameters but
cover a broad range in metallicity.
Oxygen and iron abundances were determined
from a standard one-dimension (1D) local thermodynamic
equilibrium analysis.   Systematic errors
were minimized using a homogeneous and
differential analysis of the bulge, halo,
thick disk, and thin disk stars.   For
further details refer to the parent paper
(\cite{mea08}).

The value of solar oxygen abundance also affects
the [O/Fe]-[Fe/H] relation.   Recent determinations
point towards lower values (e.g.,  \cite{apa01};
\cite{som01}; \cite{asa04};
\cite{mel04}), but the question is still
under debate (e.g., \cite{laa07}; \cite{snn07});
\cite{csn08}; \cite{caa08}).
A solar oxygen mass abundance,
$(Z_{\rm O})_\odot=0.0056$, deduced from 
\cite{apa01},
shall be used to preserve comparison
with earlier results (C01, C07, C08).

Due to the above mentioned uncertainties, it has
been preferred to deal with less recent iron abundance
determinations to preserve comparison with earlier
work (C01, C07, C08), where direct oxygen abundance
determinations are still lacking.   More recent,
statistically more significant and less well scrutinised
determinations e.g., the Geneva-Copenhagen survey
(\cite{noa04}) for the SN disk and
the Hamburg/ESO survey (\cite{sca08})
for the metal-poor
SN halo, are related to iron abundance instead of oxygen
abundance as in e.g., Ra07 and Ma08 samples, and for
this reason shall not be used in the current attempt.

The empirical, differential, oxygen abundance distribution
(EDOD), $\psi=\Delta N/$ $(N\Delta\phi)$, $\log\phi
=$[O/H], is determined in Section  \ref{s:idata}
for RaK07, RaN07, RaH07, RaU07, and MaK08, MaN08,
MaB08, subsamples.   In addition, a
putative EDOD is derived for the SN thick + thin
disk by weighting the mass of the related subsystems,
in Section \ref{s:imd}.   A comparison with the
predictions of simple inhomogeneous models, is given
in Section \ref{s:ism}.   The discussion 
and related implications for the formation of the
Galaxy are the subject of Section \ref{s:disc}.
Some concluding remarks are reported in Section \ref{s:conc}.
%

\section{The data} \label{s:idata}
\subsection{The empirical differential oxygen
abundance distribution} \label{ss:EDOD}

While observations are related to logarithmic
number abundances, [E/H]=$\log({\rm E/H})-\log({\rm E}_\odot
/{\rm H}_\odot)$, models of chemical evolution deal
with mass abundances, $Z_{\rm E}=M_{\rm E}/M$,
where E is a generic element heavier than He,
$M_{\rm E}$ and $M$ are the total mass under
the form of the element, E, and of all elements,
respectively.   The following relation (\cite{pag89};
\cite{maa93}; \cite{rom96}; C00, C01, C07, C08):
\begin{equation}
\label{eq:lgfi}
\log\phi=\log\frac {Z_{\rm O}}{(Z_{\rm O})_\odot}=
\left[\frac{\rm O}{\rm H}\right]~~;
\end{equation}
holds to a good extent.   For a formal derivation
refer to earlier work (C07, Appendix A).

Following previous attempts (\cite{pag89}; \cite{maa93}; 
\cite{rom96}), the 
comparison
between model predictions and observations shall
be performed using the differential instead of
the cumulative metallicity distribution, as it
is a more sensitive test (\cite{pag89}) and
allows direct comparison between different
samples (\cite{rom96}).   The
occurrence of a sensitivity error in performing
observations, precludes the use of a proper
differential notation in dealing with the
EDOD.   This is the reason why the bin
length cannot be lower than
the sensitivity error, and differential ratios
i.e. first derivatives, $\diff N/\diff\phi$,
must be replaced by increment ratios, $\Delta N
/\Delta\phi$, where $\Delta\phi$ is the bin
length, $\Delta N$ the number of sample objects
within the selected bin, and $N$ is the total
number of sample objects.

Accordingly, the EDOD in a selected class of
objects is defined as:
\begin{leftsubeqnarray}
\slabel{eq:fiba}
&& \psi(\phi\mp\Delta^\mp\phi)=\log\frac{\Delta N}{N\Delta
\phi}~~; \\
\slabel{eq:fibb}
&& \phi=\frac{\phi^++\phi^-}2~~;\qquad\Delta^\mp\phi=\frac{\phi^+-\phi^-}2~~;
\\
\slabel{eq:fibc}
&& \phi^\mp=\exp_{10}{\rm [O/H]}^\mp~~;
\label{seq:fib}
\end{leftsubeqnarray}
where in general, $\exp_\xi$ defines the power
of basis, $\xi$ (in particular, $\exp$
defines the power of basis, e, according to the
standard notation), and
the increment ratio, $\Delta N/
\Delta\phi$, used in earlier attempts
(\cite{pag89}; \cite{maa93}) has
been replaced by its normalized
counterpart, $\Delta N/$ $(N\Delta\phi)$,
used in more recent investigations
(\cite{rom96}; C00,
C01, C07, C08), to allow comparison between
different samples.
The uncertainty on $\Delta N$
has been evaluated from Poisson errors (e.g., 
\cite{ryn91}), as $\sigma_{\Delta N}=
(\Delta N)^{1/2}$, and the related uncertainty 
in the EDOD is (e.g., C01, C07, C08):
\begin{leftsubeqnarray}
\slabel{eq:psiera}
&& \Delta^\mp\psi=\vert\psi-\psi^\mp\vert=
\log\left[1\mp\frac{(\Delta N)^
{1/2}}{\Delta N}\right]~~; \\
\slabel{eq:psierb}
&& \psi^\mp=\log\frac{\Delta N\mp(\Delta N)^{1/2}}{N\Delta\phi}~~;
\label{seq:psier}
\end{leftsubeqnarray}
where $\psi^-\rightarrow-\infty$ in the limit
$\Delta N\rightarrow1$.   For further details
refer to earlier work (C01).

In the following tables and figures, the bin
sizes in normalized oxygen
abundance, $\phi$, shall correspond to uniform bin
sizes in [O/H] (or [Fe/H] if an empirical
[O/H]-[Fe/H] relation is used),
which implies non uniform bin sizes,
$\Delta\phi=2\Delta^\mp\phi$, as shown in
Table \ref{t:OHP}.
\begin{table}
\caption[par]{Logarithmic and numerical bins in
oxygen abundance.   Upper and lower values for
each bin are denoted as [O/H]$^\mp=\log\phi^\mp$,
according to Eqs.\,(\ref{eq:lgfi}) and (\ref
{eq:fibb}).   Logarithmic
bins which differ by 1 dex are related to
numerical bins which differ by a factor 10,
which makes an (up and down) endless recursion
of the table}
\label{t:OHP}
\begin{center}
\begin{tabular}{llll}
\multicolumn{1}{c}{[O/H]$^-$} &
\multicolumn{1}{c}{[O/H]$^+$} &
\multicolumn{1}{c}{$\phi$} &
\multicolumn{1}{c}{$\Delta^\mp\phi$} \\
\noalign{\smallskip}
\hline\noalign{\smallskip}
0.05 & 0.15 & 1.2673~E+0 & 1.4526~E$-$1 \\
0.15 & 0.25 & 1.5954~E+0 & 1.8287~E$-$1 \\
0.25 & 0.35 & 2.0085~E+0 & 2.3022~E$-$1 \\
0.35 & 0.45 & 2.5286~E+0 & 2.8983~E$-$1 \\
0.45 & 0.55 & 3.1833~E+0 & 3.6488~E$-$1 \\
0.55 & 0.65 & 4.0075~E+0 & 4.5935~E$-$1 \\
0.65 & 0.75 & 5.0451~E+0 & 5.7829~E$-$1 \\
0.75 & 0.85 & 6.3514~E+0 & 7.2802~E$-$1 \\
0.85 & 0.95 & 9.9960~E+0 & 9.1653~E$-$1 \\
0.95 & 1.05 & 1.0066~E+1 & 1.1538~E$-$0 \\
1.05 & 1.15 & 1.2673~E+1 & 1.4526~E$-$0 \\
\noalign{\smallskip}
\hline
\end{tabular}
\end{center}
\end{table}

\subsection{Oxygen abundance distribution in
subsamples} \label{ss:KN}

The Ra07 sample is considered, where oxygen
abundance is directly inferred both in
presence and in absence of the local thermodynamic
equilibrium (LTE) approximation (\cite{raa07},
Table 6, available at the CDS).   The LTE
approximation well holds deep in the stellar
photosphere, and a continuous transition occurs
up to complete non-equilibrum (NLTE) high in
the atmosphere.   For further details refer to
specialized textbooks (e.g., \cite{gra05}, Chapter 6).

The metallicity range is $-1.5\le{\rm[Fe/H]}
\le0.5$, but the sample is biased against lower
metallicities, [Fe/H]$\appleq-1$.   On the other
hand, it is expected to be unbiased at higher
metallicities, for the following reasons: (i)
more than 90\% of the sample stars are located
within 100 pc from the sun, and (ii) the related
differential iron abundance distribution looks
similar to its counterpart deduced from different
SN disk samples (\cite{wyg95}; \cite{jor00};
\cite{hay01}; \cite{fuh08}), and the same
holds for SN thick disk and thin disk subsamples,
when comparison may be performed.   The EDOD has
been deduced from the Ra07 sample in an earlier
attempt (C08), both in presence and in absence
of LTE approximation.

Under the assumption that sample objects belong
to three stellar populations (thick disk, thin
disk, halo), each with a Gaussian velocity
distribution, the probability of a star being
located in one of the components, may be
expressed as a function of the Galactic space
velocities, $P_i=P_i(U,V,W)$, where $i=1,2,3,$
for the thin disk, the thick disk, and the halo,
respectively; a $P_i>0.7$ constraint is set
to determine the membership of sample objects
to the $i$ population (\cite{raa07}).

Using the above mentioned prescriptions, the
Ra07 sample has been subdivided into four
subsamples: RaN07, RaK07, RaH07, RaU07, where
$P_1>0.7$, $P_2>0.7$, $P_3>0.7$, $P_i\le0.7$,
$i=1,2,3,$ respectively.   The RaH07 subsample
is by far incomplete, as [Fe/H]$>-1.5$ for
sample objects and any attempt to trace a halo
abundance trend using only the metal-rich tail
is dubious (Ramirez et al., 2007).   With this
caveat in mind, the RaH07 subsample shall be
considered together with the RaU07 subsample.

On the other hand, it can be seen (\cite{raa07})
that a
similar [O/Fe]-[Fe/H] empirical relation
is exhibited by RaK07 and RaH07 subsamples:
\begin{equation}
\label{eq:OFe}
{\rm [O/Fe]}=a_{\rm O}+b_{\rm O}{\rm [Fe/H]}~~;
\end{equation}
where $a_{\rm O}=0.370\mp0.027,$ $b_{\rm O}=-0.121\mp
0.043,$  $N_{\rm O}=43,$ $\sigma_{\rm O}=0.065,$ for
the RaK07 subsample; $a_{\rm O}=0.388\mp0.049,$
$b_{\rm O}=-0.048\mp0.071,$  $N_{\rm O}=12,$ $\sigma_{\rm O}=
0.072,$ for the RaH07 subsample; in
addition, $N_{\rm O}$ is the number of stars
used for the fit and $\sigma_{\rm O}$ the random
scatter around the mean fit.

The Ma08 sample is also considered,
where oxygen abundance is deduced
from a standard 1D local thermodynamic
equilibrium analysis, and systematic
errors are minimized using a homogeneous
and differential analysis of stars
belonging to different populations
(\cite{mea08}).   The Ma08
sample is made of four subsamples:
MaK08, MaN08, MaH08, MaB08, within
the metallicity range, $-1.7\le{\rm
[Fe/H]}\le0.5$.

A similar fit, expressed by Eq.\,(\ref
{eq:OFe}), is shown using MaB08 and MaK08
subsamples, where $a_{\rm O}=0.41,$
$b_{\rm O}=-0.02,$  $N_{\rm O}=19,$
$\sigma_{\rm O}=0.09,$ for the
MaB08 subsample, and $a_{\rm O}=0.39,$
$b_{\rm O}=-0.01,$  $N_{\rm O}=21,$
$\sigma_{\rm O}=0.09,$ for the MaK08
subsample (\cite{mea08}).
It can also be seen that
oxygen abundance patterns exhibited
by the MaH08 subsample are very close
to their counterparts related to the
MaB08 and MaK08 subsamples (\cite{mea08},
Fig.\,2).

In conclusion, a similar trend for
the [O/Fe]-[Fe/H] relation with
respect to the SN thick disk, the bulge,
and the SN halo, is deduced from both Ra07 and
Ma08 samples, in the metallicity range
where the thick disk is unambiguously
defined.

\subsection{Oxygen abundance distribution
in SN thick disk} \label{ss:KD}

Both RaK07 and MaK08 subsamples are considered
to infer separate EDODs related to the SN thick
disk, for comparison with model predictions.
The oxygen abundance range is $-0.60<{\rm [O/H]}
<0.45$ (LTE), $-0.80\le{\rm [O/H]}<0.45$ (NLTE),
for the RaK07 subsample, and $-1.40<{\rm [O/H]}<0.30$
for the MaK08 subsample.

\begin{figure*}[t]
\begin{center}
\includegraphics[scale=0.8]{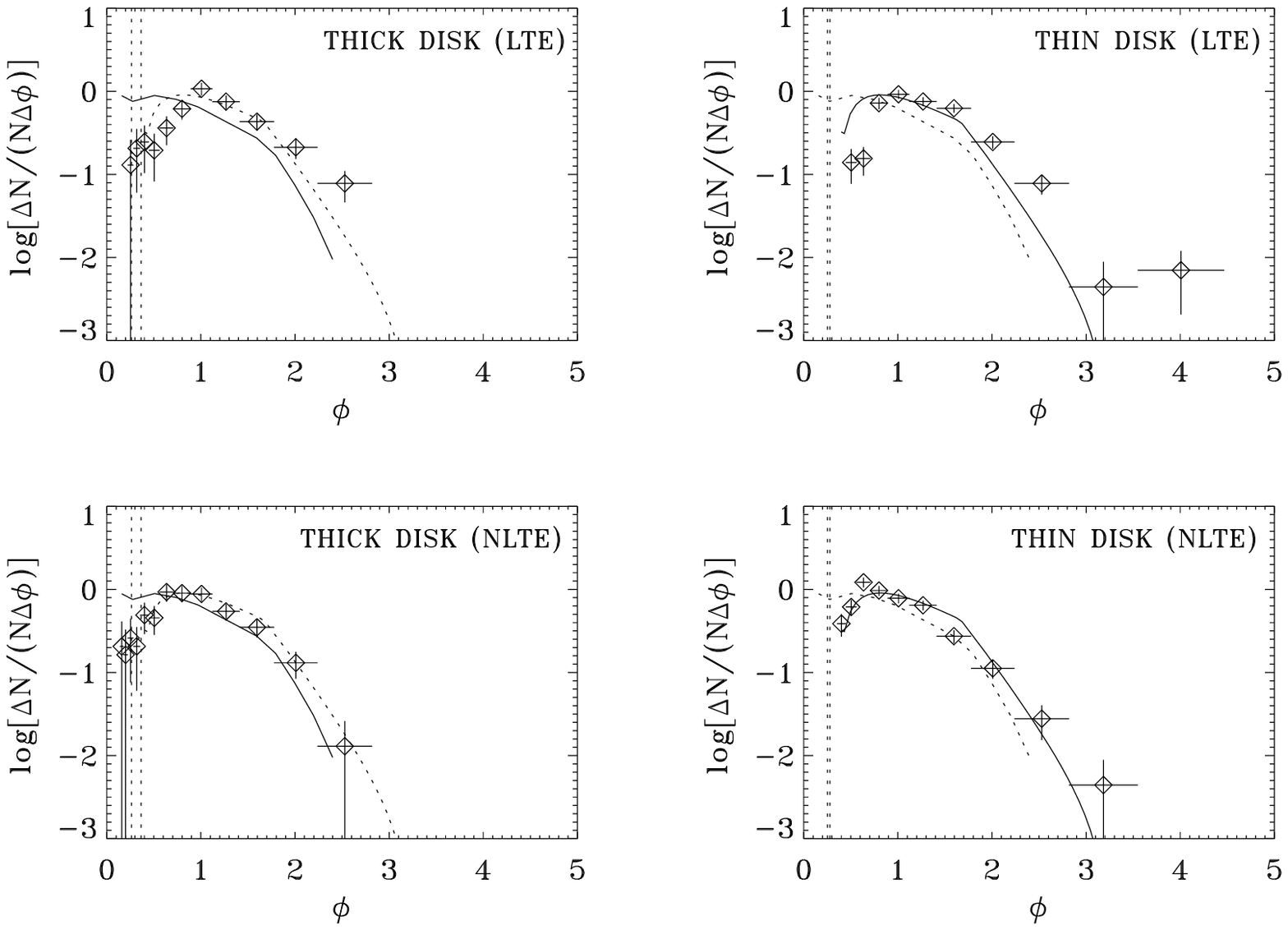}
\caption[OKND]{The empirical, differential oxygen abundance
distribution (EDOD) related to the RaK07 (left panels) and
RaN07 (right panels) subsample, both in presence (LTE, top panels)
and in absence (NLTE, bottom panels) of the local thermodynamic
equilibrium approximation.   The theoretical, differential
oxygen abundance distribution (TDOD) related to inhomogeneous
models of chemical evolution for the solar neighbourhood (SN)
thick disk (case DB15, left panels) and the SN thin disk (case DN,
right panels) are plotted as full curves, and as dotted curves
in the remaining panels.   Models are fitted to NLTE data.
The dashed vertical bands correspond to [Fe/H]$=-1$ and related
uncertainties, deduced from Eq.\,(\ref{eq:OFe}) in dealing
with the thick (left panels) and the thin (right panels)
disk (\cite{raa07}}
\label{f:OKND}
\end{center}
\end{figure*}

The EDOD related to the RaK07 subsample is listed
in Table \ref{t:RaK07} and plotted in 
Fig.\,\ref{f:OKND}, upper (LTE) and
lower (NLTE) left, where the dashed
vertical bands correspond to [Fe/H]$=
-1$ and related uncertainty, deduced
from Eq.\,(\ref{eq:OFe}) in dealing
with the thick disk (\cite{raa07}).
\begin{table*}
\caption[par]{The empirical, differential
oxygen abundance distribution (EDOD) in the
solar neighbourhood
(SN) thick disk, deduced from the RaK07 subsample
($N=133$) both in presence (LTE) and in 
absence (NLTE) of the local thermodynamic
equilibrium approximation}
\label{t:RaK07}
\begin{center}
\begin{tabular}{rrrrrrrrr}
\multicolumn{1}{c|}{\phantom{$\phi$}}
&\multicolumn{4}{c|}{LTE}
&\multicolumn{4}{c|}{NLTE} \\
\hline\noalign{\smallskip}
\multicolumn{1}{c}{$\phi$} & \multicolumn{1}{c}{$\phantom{0}\psi$} &
\multicolumn{1}{c}{$\Delta^-\psi$} & \multicolumn{1}{c}{$\Delta^+\psi$} &
\multicolumn{1}{c}{$\Delta N$}  &
\multicolumn{1}{c}{$\phantom{0}\psi$} &
\multicolumn{1}{c}{$\Delta^-\psi$} & \multicolumn{1}{c}{$\Delta^+\psi$} &
\multicolumn{1}{c}{$\Delta N$}  \\
\noalign{\smallskip}
\hline\noalign{\smallskip}
1.595~E$-$1 &                &             &             &    & $-$6.870~E$-$1 & $+\infty$   & 3.010~E$-$1 & 1  \\
2.009~E$-$1 &                &             &             &    & $-$7.870~E$-$1 & $+\infty$   & 3.010~E$-$1 & 1  \\
2.529~E$-$1 & $-$8.870~E$-$1 & $+\infty$   & 3.010~E$-$1 & 1  & $-$5.860~E$-$1 & 5.333~E$-$1 & 2.323~E$-$1 & 2  \\
3.183~E$-$1 & $-$6.860~E$-$1 & 5.333~E$-$1 & 2.323~E$-$1 & 2  & $-$6.860~E$-$1 & 5.333~E$-$1 & 2.323~E$-$1 & 2  \\
4.008~E$-$1 & $-$6.099~E$-$1 & 3.740~E$-$1 & 1.979~E$-$1 & 3  & $-$3.089~E$-$1 & 2.279~E$-$1 & 1.487~E$-$1 & 6  \\
5.045~E$-$1 & $-$7.099~E$-$1 & 3.740~E$-$1 & 1.979~E$-$1 & 3  & $-$3.419~E$-$1 & 2.062~E$-$1 & 1.392~E$-$1 & 7  \\
6.351~E$-$1 & $-$4.419~E$-$1 & 2.062~E$-$1 & 1.392~E$-$1 & 7  & $-$3.175~E$-$2 & 1.167~E$-$1 & 9.191~E$-$2 & 18 \\
7.996~E$-$1 & $-$2.109~E$-$1 & 1.297~E$-$1 & 9.975~E$-$2 & 15 & $-$4.460~E$-$2 & 1.041~E$-$1 & 8.393~E$-$2 & 22 \\
1.007~E$-$0 & $+$3.149~E$-$2 & 8.306~E$-$2 & 6.970~E$-$2 & 33 & $-$5.566~E$-$2 & 9.283~E$-$2 & 7.644~E$-$2 & 27 \\
1.267~E$-$0 & $-$1.246~E$-$1 & 8.921~E$-$2 & 7.397~E$-$2 & 29 & $-$2.648~E$-$1 & 1.069~E$-$1 & 8.572~E$-$2 & 21 \\
1.595~E$-$0 & $-$3.648~E$-$1 & 1.069~E$-$1 & 8.572~E$-$2 & 21 & $-$4.566~E$-$1 & 1.206~E$-$1 & 9.431~E$-$2 & 17 \\
2.009~E$-$0 & $-$6.731~E$-$1 & 1.411~E$-$1 & 1.063~E$-$1 & 13 & $-$8.833~E$-$1 & 1.895~E$-$1 & 1.315~E$-$1 & 8  \\
2.529~E$-$0 & $-$1.109~E$-$0 & 1.487~E$-$1 & 1.487~E$-$1 & 6  & $-$1.887~E$-$0 & $+\infty$   & 3.010~E$-$1 & 1  \\
\noalign{\smallskip}
\hline
\end{tabular}
\end{center}
\end{table*}

\begin{figure*}[t]
\begin{center}
\includegraphics[scale=0.8]{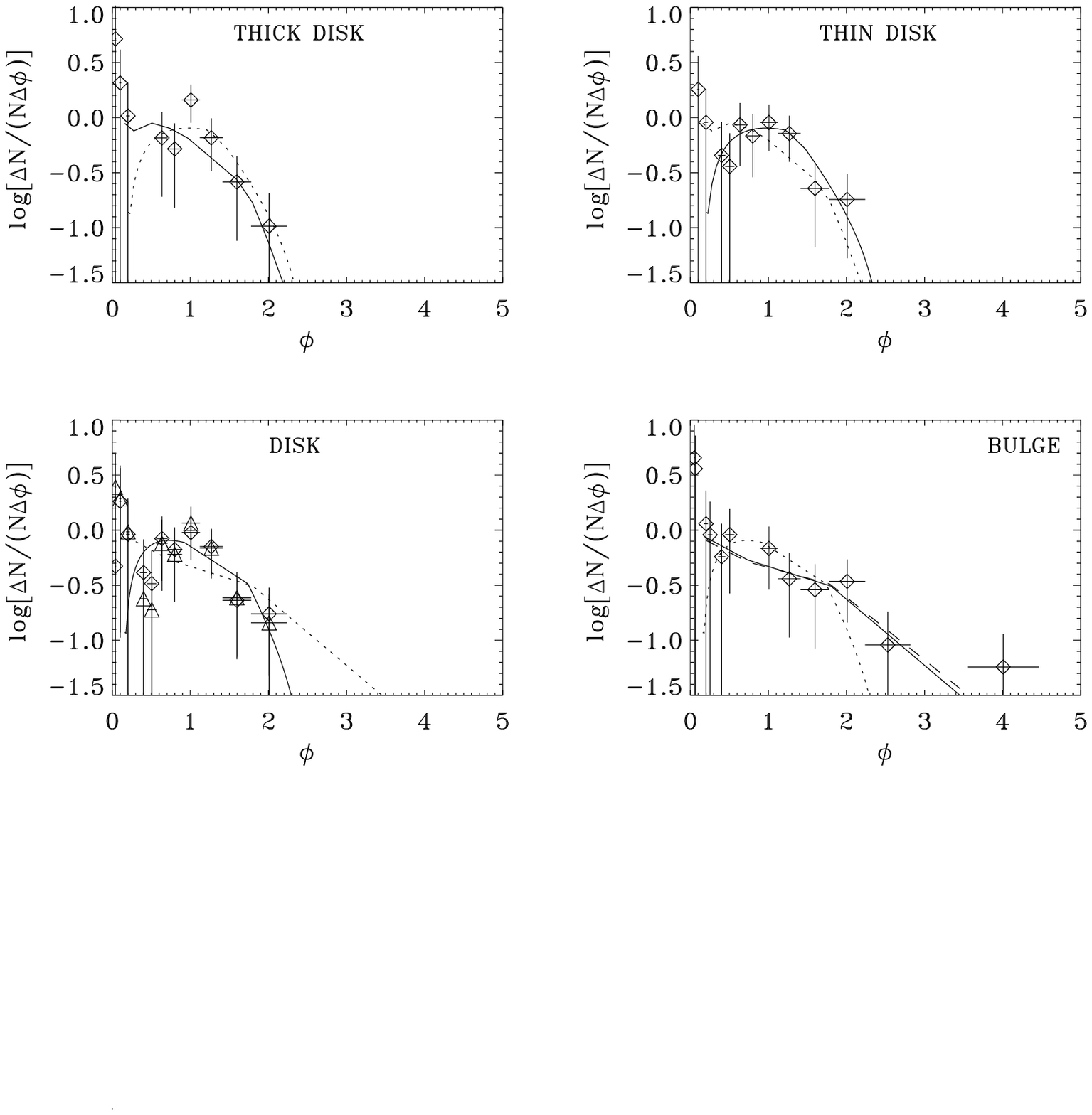}
\caption[TDBA]{The empirical, differential oxygen abundance
distribution (EDOD) related to the (from top left in
clockwise sense) MaK08, MaN08, MaB08, subsample, and a
combination of MaK08 and MaN08 subsamples after weighting
by mass, for an assumed solar neighbourhood (SN) thick
disk to thin disk mass ratio, $M_{\rm KD}/M_{\rm ND}=0.1$
(diamonds) and 0.9 (triangles).   The theoretical, differential
oxygen abundance distribution (TDOD) related to inhomogeneous
models of chemical evolution for the SN thick disk (case DB15), the
SN thin disk (case DN81), the SN thick + thin disk (case KN4), and the
bulge (case B) are plotted as full curves.   The dotted curve on
each panel reproduces (for comparison) the full curve on the
other panel of the same row.   The dashed curve plotted on
the bottom right panel corresponds to a different bulge model
(case B1)}
\label{f:TDBA}
\end{center}
\end{figure*}

The EDOD related to the MaK08 subsample is listed
in Table \ref{t:MaD08} (left) and plotted in 
Fig.\,\ref{f:TDBA} (upper left).
\begin{table*}
\caption[par]{The empirical, differential
oxygen abundance distribution (EDOD) in the
solar neighbourhood (SN) thick disk (left)
and SN thin disk (right), deduced from the
MaN08 ($N=24$) and the MaK08 ($N=21$) subsample,
respectively}
\label{t:MaD08}
\begin{center}
\begin{tabular}{rrrrrrrrr}
\multicolumn{1}{c|}{\phantom{$\phi$}}
&\multicolumn{4}{c|}{THICK DISK}
&\multicolumn{4}{c|}{THIN DISK} \\
\hline\noalign{\smallskip}
\multicolumn{1}{c}{$\phi$} & \multicolumn{1}{c}{$\phantom{0}\psi$} &
\multicolumn{1}{c}{$\Delta^-\psi$} & \multicolumn{1}{c}{$\Delta^+\psi$} &
\multicolumn{1}{c}{$\Delta N$}  &
\multicolumn{1}{c}{$\phantom{0}\psi$} &
\multicolumn{1}{c}{$\Delta^-\psi$} & \multicolumn{1}{c}{$\Delta^+\psi$} &
\multicolumn{1}{c}{$\Delta N$}  \\
\noalign{\smallskip}
\hline\noalign{\smallskip}
4.008~E$-$2 & $+$7.146~E$-$1 & $+\infty$   & 3.010~E$-$1 & 1 &                &             &             &   \\
5.045~E$-$2 &                &             &             &   &                &             &             &   \\
6.351~E$-$2 &                &             &             &   &                &             &             &   \\
7.996~E$-$2 &                &             &             &   &                &             &             &   \\
1.007~E$-$1 & $+$3.146~E$-$1 & $+\infty$   & 3.010~E$-$1 & 1 & $+$2.566~E$-$1 & $+\infty$   & 3.010~E$-$1 & 1 \\
1.267~E$-$1 &                &             &             &   &                &             &             &   \\
1.595~E$-$1 &                &             &             &   &                &             &             &   \\
2.009~E$-$1 & $+$1.461~E$-$2 & $+\infty$   & 3.010~E$-$1 & 1 & $-$4.339~E$-$2 & $+\infty$   & 3.010~E$-$1 & 1 \\
2.529~E$-$1 &                &             &             &   &                &             &             &   \\
3.183~E$-$1 &                &             &             &   &                &             &             &   \\
4.008~E$-$1 &                &             &             &   & $-$3.4343E$-$1 & $+\infty$   & 3.010~E$-$1 & 1 \\
5.045~E$-$1 &                &             &             &   & $-$4.434~E$-$1 & $+\infty$   & 3.010~E$-$1 & 1 \\
6.351~E$-$1 & $-$1.844~E$-$1 & 5.333~E$-$1 & 2.323~E$-$1 & 2 & $-$6.627~E$-$2 & 3.740~E$-$1 & 1.979~E$-$1 & 3 \\
7.996~E$-$1 & $-$2.844~E$-$1 & 5.333~E$-$1 & 2.323~E$-$1 & 2 & $-$1.663~E$-$1 & 3.740~E$-$1 & 1.979~E$-$1 & 3 \\
1.007~E$-$0 & $+$1.597~E$-$1 & 2.062~E$-$1 & 1.392~E$-$1 & 7 & $-$4.442~E$-$2 & 2.574~E$-$1 & 1.605~E$-$1 & 5 \\
1.267~E$-$0 & $-$1.833~E$-$1 & 3.010~E$-$1 & 1.761~E$-$1 & 4 & $-$1.444~E$-$1 & 2.574~E$-$1 & 1.605~E$-$1 & 5 \\
1.595~E$-$0 & $-$5.844~E$-$1 & 5.333~E$-$1 & 2.323~E$-$1 & 2 & $-$6.424~E$-$1 & 5.333~E$-$1 & 2.323~E$-$1 & 2 \\
2.009~E$-$0 & $-$9.854~E$-$1 & $+\infty$   & 3.010~E$-$1 & 1 & $-$7.424~E$-$1 & 5.333~E$-$1 & 2.323~E$-$1 & 2 \\
\noalign{\smallskip}
\hline
\end{tabular}
\end{center}
\end{table*}

It can be seen that EDODs deduced from RaK07 and MaK08
subsamples show qualitative agreement, leaving aside
the low-metallicity ([Fe/H]$<-1$) tail where, on the
other hand, oxygen abundance bins are poorly populated
and the statistical relevance is weak.

\subsection{Oxygen abundance distribution
in SN thin disk} \label{ss:ND}

Both RaN07 and MaN08 subsamples are considered
to infer separate EDODs related to the SN thin
disk, for comparison with model predictions.
The oxygen abundance range is $-0.35<{\rm [O/H]}
<0.60$ (LTE), $-0.45<{\rm [O/H]}<0.45$ (NLTE),
for the RaN07 subsample, and $-1.00<{\rm [O/H]}<0.30$
for the MaN08 subsample.

The EDOD related to the RaN07 subsample is listed
in Table \ref{t:RaN07} and plotted in 
Fig.\,\ref{f:OKND}, upper (LTE) and
lower (NLTE) right, where the dashed
vertical bands correspond to [Fe/H]$=
-1$ and related uncertainty, deduced
from Eq.\,(\ref{eq:OFe}) in dealing
with the SN thin disk (\cite{raa07}).

\begin{table*}
\caption[par]{The empirical, differential
oxygen abundance distribution (EDOD) in the
solar neighbourhood
(SN) thin disk, deduced from the RaN07 subsample
($N=310$) both in presence (LTE) and in 
absence (NLTE) of the local thermodynamic
equilibrium approximation}
\label{t:RaN07}
\begin{center}
\begin{tabular}{rrrrrrrrr}
\multicolumn{1}{c|}{\phantom{$\phi$}}
&\multicolumn{4}{c|}{LTE}
&\multicolumn{4}{c|}{NLTE} \\
\hline\noalign{\smallskip}
\multicolumn{1}{c}{$\phi$} & \multicolumn{1}{c}{$\phantom{0}\psi$} &
\multicolumn{1}{c}{$\Delta^-\psi$} & \multicolumn{1}{c}{$\Delta^+\psi$} &
\multicolumn{1}{c}{$\Delta N$}  &
\multicolumn{1}{c}{$\phantom{0}\psi$} &
\multicolumn{1}{c}{$\Delta^-\psi$} & \multicolumn{1}{c}{$\Delta^+\psi$} &
\multicolumn{1}{c}{$\Delta N$}  \\
\noalign{\smallskip}
\hline\noalign{\smallskip}
4.008~E$-$1 &                &             &             &    & $-$4.131~E$-$1 & 1.558~E$-$1 & 1.145~E$-$1 & 11 \\
5.045~E$-$1 & $-$8.556~E$-$1 & 2.574~E$-$1 & 1.605~E$-$1 & 5  & $-$2.121~E$-$1 & 1.041~E$-$1 & 8.393~E$-$2 & 22 \\
6.351~E$-$1 & $-$8.094~E$-$1 & 2.062~E$-$1 & 1.392~E$-$1 & 7  & $+$8.583~E$-$2 & 6.290~E$-$2 & 5.494~E$-$2 & 55 \\
7.996~E$-$1 & $-$1.418~E$-$1 & 7.375~E$-$2 & 6.302~E$-$2 & 41 & $-$1.417~E$-$2 & 6.290~E$-$2 & 5.494~E$-$2 & 55 \\
1.007~E$-$0 & $-$3.499~E$-$2 & 5.704~E$-$2 & 5.042~E$-$2 & 66 & $-$1.064~E$-$1 & 6.230~E$-$2 & 5.447~E$-$2 & 56 \\
1.267~E$-$0 & $-$1.220~E$-$1 & 5.614~E$-$2 & 4.971~E$-$2 & 68 & $-$1.911~E$-$1 & 6.113~E$-$2 & 5.358~E$-$2 & 58 \\
1.595~E$-$0 & $-$2.033~E$-$1 & 5.487~E$-$2 & 3.871~E$-$2 & 71 & $-$5.632~E$-$1 & 8.598~E$-$2 & 7.174~E$-$2 & 31 \\
2.009~E$-$0 & $-$6.105~E$-$1 & 8.042~E$-$2 & 6.783~E$-$2 & 35 & $-$9.504~E$-$1 & 1.249~E$-$1 & 9.691~E$-$2 & 16 \\
2.529~E$-$0 & $-$1.108~E$-$0 & 1.351~E$-$1 & 1.029~E$-$1 & 14 & $-$1.556~E$-$0 & 2.574~E$-$1 & 1.605~E$-$1 & 5  \\
3.183~E$-$0 & $-$2.355~E$-$0 & $+\infty$   & 3.010~E$-$1 & 1  & $-$2.355~E$-$0 & $+\infty$   & 3.010~E$-$1 & 1  \\
4.008~E$-$0 & $-$2.154~E$-$0 & 5.333~E$-$1 & 2.323~E$-$1 & 2  &                &             &             &    \\
\noalign{\smallskip}
\hline
\end{tabular}
\end{center}
\end{table*}

The EDOD related to the MaN08 subsample is listed
in Table \ref{t:MaD08} (right) and plotted in 
Fig.\,\ref{f:TDBA} (upper right).

It can be seen that EDODs deduced from RaN07 and MaN08
subsamples show qualitative agreement, leaving aside
the low-metallicity ([Fe/H]$<-1$) tail where, on the
other hand, oxygen abundance bins are poorly populated
and the statistical relevance is weak.

\subsection{Oxygen abundance distribution in the bulge}
\label{ss:B}

The MaB08 subsample is considered
to infer the EDOD related to the bulge,
for comparison with model predictions.
The oxygen abundance range is $-1.3<{\rm [O/H]}
<0.5$.

The EDOD related to the MaB08 subsample is listed
in Table \ref{t:MaB08} and plotted in 
Fig.\,\ref{f:TDBA} (lower right).
\begin{table}
\caption[par]{The empirical, differential
oxygen abundance distribution (EDOD) in the
bulge, deduced from the MaB08 ($N=19$) subsample}
\label{t:MaB08}
\begin{center}
\begin{tabular}{rrrrr}
\multicolumn{1}{c|}{\phantom{$\phi$}}
&\multicolumn{4}{c|}{BULGE} \\
\hline\noalign{\smallskip}
\multicolumn{1}{c}{$\phi$} & \multicolumn{1}{c}{$\phantom{0}\psi$} &
\multicolumn{1}{c}{$\Delta^-\psi$} & \multicolumn{1}{c}{$\Delta^+\psi$} &
\multicolumn{1}{c}{$\Delta N$}  \\
\noalign{\smallskip}
\hline\noalign{\smallskip}
5.045~E$-$2 & $+$6.581~E$-$1 & $+\infty$   & 3.010~E$-$1 & 1 \\
6.351~E$-$2 & $+$5.581~E$-$1 & $+\infty$   & 3.010~E$-$1 & 1 \\
7.996~E$-$2 &                &             &             &   \\
1.007~E$-$1 &                &             &             &   \\
1.267~E$-$1 &                &             &             &   \\
1.595~E$-$1 &                &             &             &   \\
2.009~E$-$1 & $+$5.807~E$-$2 & $+\infty$   & 3.010~E$-$1 & 1 \\
2.529~E$-$1 & $-$4.193~E$-$2 & $+\infty$   & 3.010~E$-$1 & 1 \\
3.183~E$-$1 &                &             &             &   \\
4.008~E$-$1 & $-$2.419~E$-$1 & $+\infty$   & 3.010~E$-$1 & 1 \\
5.045~E$-$1 & $-$4.090~E$-$2 & 5.333~E$-$1 & 2.323~E$-$1 & 2 \\
6.351~E$-$1 &                &             &             &   \\
7.996~E$-$1 &                &             &             &   \\
1.007~E$-$0 & $-$1.648~E$-$1 & 3.740~E$-$1 & 1.979~E$-$1 & 3 \\
1.267~E$-$0 & $-$4.409~E$-$1 & 5.333~E$-$1 & 2.323~E$-$1 & 2 \\
1.595~E$-$0 & $-$5.409~E$-$1 & 5.333~E$-$1 & 2.323~E$-$1 & 2 \\
2.009~E$-$0 & $-$4.648~E$-$1 & 3.740~E$-$1 & 1.579~E$-$1 & 3 \\
2.529~E$-$0 & $-$1.042~E$-$0 & $+\infty$   & 3.010~E$-$1 & 1 \\
3.183~E$-$0 &                &             &             &   \\
4.008~E$-$0 & $-$1.242~E$-$0 & $+\infty$   & 3.010~E$-$1 & 1 \\
\noalign{\smallskip}
\hline
\end{tabular}
\end{center}
\end{table}

It can be seen that EDODs deduced from the MaN08
subsample and the Sa96 sample, where [O/H]-[Fe/H]
empirical relations have been used (C07), show
qualitative agreement, in spite of the
poorly populated oxygen abundance bins, for
which the statistical relevance is weak, in the
former case.

\subsection{Oxygen abundance distribution in the SN halo}
\label{ss:H}

The RaH07 subsample is considered to infer the EDOD
related to the SN halo only for illustrative purposes,
due to its incompleteness.   The oxygen abundance
range is $-1.00<{\rm [O/H]}<0.30$ (LTE), $-1.20<{\rm [O/H]}
<0.20$ (NLTE).

\begin{figure*}[t]
\begin{center}
\includegraphics[scale=0.8]{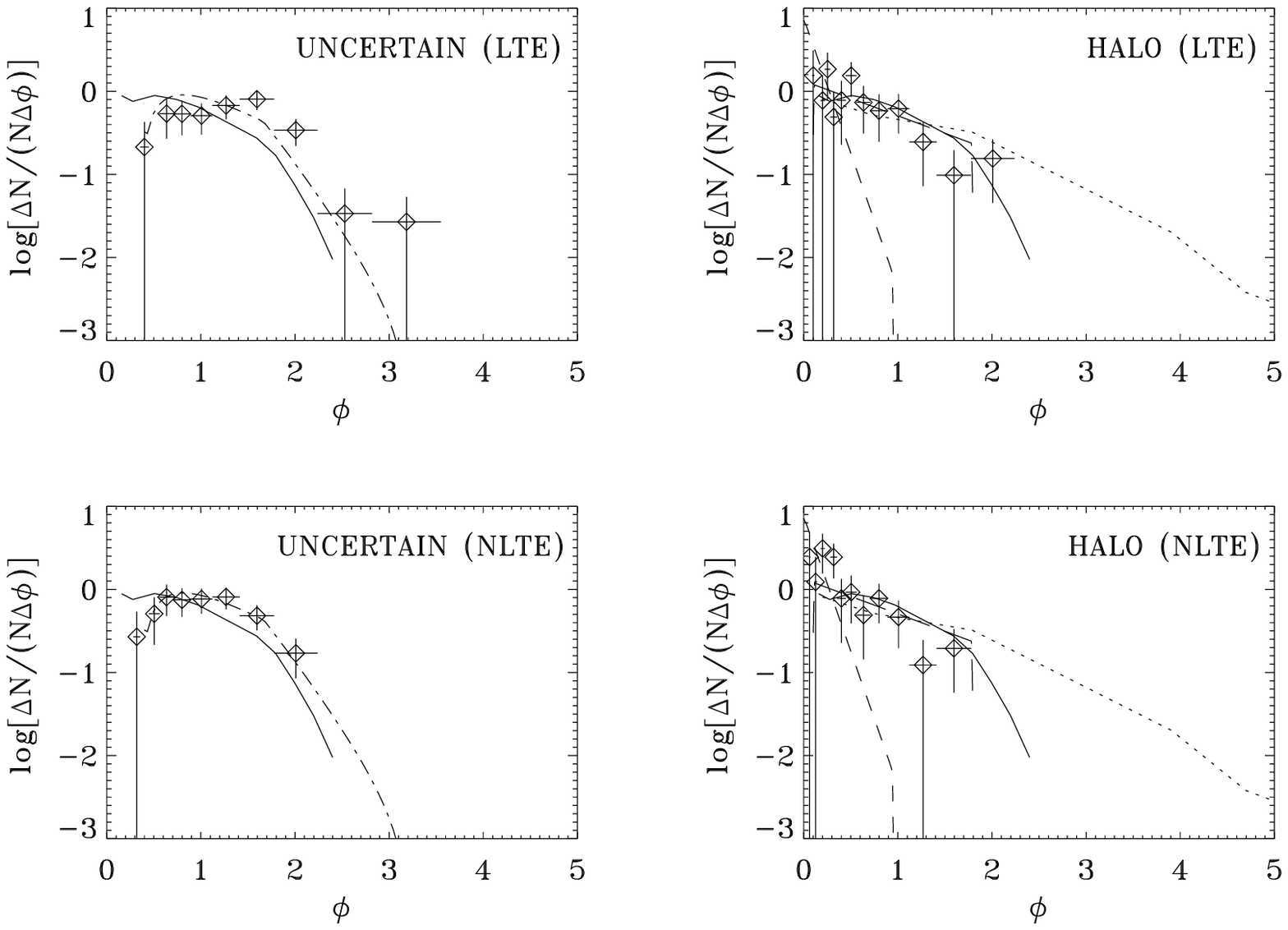}
\caption[THUD]{The empirical, differential oxygen abundance
distribution (EDOD) related to the RaU07 (left panels) and
RaH07 (right panels) subsample, both in presence (LTE, top panels)
and in absence (NLTE, bottom panels) of the local thermodynamic
equilibrium approximation.   The theoretical, differential
oxygen abundance distribution (TDOD) related to inhomogeneous
models of chemical evolution for the solar neighbourhood (SN)
thick disk (case DB15), the SN thin disk (case DN), and the SN halo
(case H)
are plotted as full curves, dot-dashed curves, and long-dashed
curves, respectively.   Models are fitted to NLTE data.  Bulge
(case B1) and SN halo (case H1) counterparts presented in an earlier
attempt (C07) are also shown as dotted and dashed curves,
respectively.
}
\label{f:THUD}
\end{center}
\end{figure*}

The EDOD related to the RaH07 subsample is listed
in Table \ref{t:RaH07} and plotted in 
Fig.\,\ref{f:THUD} both in presence (upper right)
and in absence (lower right) of the LTE approximation.
\begin{table*}
\caption[par]{The empirical, differential
oxygen abundance distribution (EDOD) in the
solar neighbourhood
(SN) halo, deduced from the RaH07 subsample
($N=28$) both in presence (LTE) and in 
absence (NLTE) of the local thermodynamic
equilibrium approximation}
\label{t:RaH07}
\begin{center}
\begin{tabular}{rrrrrrrrr}
\multicolumn{1}{c|}{\phantom{$\phi$}}
&\multicolumn{4}{c|}{LTE}
&\multicolumn{4}{c|}{NLTE} \\
\hline\noalign{\smallskip}
\multicolumn{1}{c}{$\phi$} & \multicolumn{1}{c}{$\phantom{0}\psi$} &
\multicolumn{1}{c}{$\Delta^-\psi$} & \multicolumn{1}{c}{$\Delta^+\psi$} &
\multicolumn{1}{c}{$\Delta N$}  &
\multicolumn{1}{c}{$\phantom{0}\psi$} &
\multicolumn{1}{c}{$\Delta^-\psi$} & \multicolumn{1}{c}{$\Delta^+\psi$} &
\multicolumn{1}{c}{$\Delta N$}  \\
\noalign{\smallskip}
\hline\noalign{\smallskip}
6.351~E$-$2 &                &             &             &   & $+$3.897~E$-$1 & $+\infty$   & 3.010~E$-$1 & 1 \\
7.996~E$-$2 &                &             &             &   &                &             &             &   \\
1.007~E$-$1 & $-$1.897~E$-$1 & $+\infty$   & 3.010~E$-$1 & 1 &                &             &             &   \\
1.267~E$-$1 &                &             &             &   & $+$8.967~E$-$2 & $+\infty$   & 3.010~E$-$1 & 1 \\
1.595~E$-$1 &                &             &             &   &                &             &             &   \\
2.009~E$-$1 & $-$1.103~E$-$1 & $+\infty$   & 3.010~E$-$1 & 1 & $+$4.917~E$-$1 & 3.010~E$-$1 & 1.761~E$-$1 & 4 \\
2.529~E$-$1 & $+$2.668~E$-$1 & 3.740~E$-$1 & 1.979~E$-$1 & 3 &                &             &             &   \\
3.183~E$-$1 & $-$3.103~E$-$1 & $+\infty$   & 3.010~E$-$1 & 1 & $+$3.886~E$-$1 & 2.574~E$-$1 & 1.605~E$-$1 & 5 \\
4.008~E$-$1 & $-$1.093~E$-$1 & 5.333~E$-$1 & 2.323~E$-$1 & 2 & $-$1.093~E$-$1 & 5.333~E$-$1 & 2.323~E$-$1 & 2 \\
5.045~E$-$1 & $+$1.886~E$-$1 & 2.574~E$-$1 & 1.605~E$-$1 & 5 & $-$3.321~E$-$2 & 3.740~E$-$1 & 1.979~E$-$1 & 3 \\
6.351~E$-$1 & $-$1.332~E$-$1 & 3.740~E$-$1 & 1.979~E$-$1 & 3 & $-$3.093~E$-$1 & 5.333~E$-$1 & 2.323~E$-$1 & 2 \\
7.996~E$-$1 & $-$2.332~E$-$1 & 3.740~E$-$1 & 1.979~E$-$1 & 3 & $-$1.083~E$-$1 & 3.010~E$-$1 & 1.761~E$-$1 & 4 \\
1.007~E$-$0 & $-$2.083~E$-$1 & 3.010~E$-$1 & 1.761~E$-$1 & 4 & $-$3.332~E$-$1 & 3.740~E$-$1 & 1.979~E$-$1 & 3 \\
1.267~E$-$0 & $-$6.093~E$-$1 & 5.333~E$-$1 & 2.323~E$-$1 & 2 & $-$9.103~E$-$1 & $+\infty$   & 3.010~E$-$1 & 1 \\
1.595~E$-$0 & $-$1.010~E$-$0 & $+\infty$   & 3.010~E$-$1 & 1 & $-$7.093~E$-$1 & 5.333~E$-$1 & 2.323~E$-$1 & 2 \\
2.009~E$-$0 & $-$8.093~E$-$1 & 5.333~E$-$1 & 2.323~E$-$1 & 2 &                &             &             &   \\
\noalign{\smallskip}
\hline
\end{tabular}
\end{center}
\end{table*}

It can be seen that EDODs deduced from RaH07
and MaB08 subsamples show qualitative agreement.

The EDOD related to the RaU07 subsample is
also plotted in Fig.\,\ref{f:THUD} for
comparison, both in presence (upper left)
and in absence (lower left) of the LTE
approximation.   The related trend appears
to be closer to its SN thick disk and SN thin
disk counterparts, with respect to the
SN halo.

\section{Inferred SN thick + thin disk
oxygen abundance distribution}
\label{s:imd}

Under the assumption of a universal
initial mass function (IMF) for a star
generation, the SN thick + thin disk
EDOD may be expressed as (C07, C08):
\begin{leftsubeqnarray}
\slabel{eq:psisa}
&& \psi=\log\left[\frac{M_{\rm KD}}{M_{\rm D}}\frac
{\Delta N_{\rm KD}}{N_{\rm KD}\Delta\phi}+
\frac{M_{\rm ND}}{M_{\rm D}}\frac{\Delta N_{\rm ND}}
{N_{\rm ND}\Delta\phi}\right]~~; \\
\slabel{eq:psisb}
&& \Delta^\mp\psi=\left\vert\log\left[1\mp
\frac{\sigma_{\Delta{\rm N}/{\rm N}}}{\Delta N/N}\right]
\right\vert~~; \\
\slabel{eq:psisc}
&& \psi^\mp=\log\left[\frac{M_{\rm KD}}{M_{\rm D}}\frac
{\Delta N_{\rm KD}\mp(\Delta N_{\rm KD})^{1/2}}{N_{\rm KD}\Delta\phi}\right.
\nonumber \\
&& \phantom{\psi^\mp=\log\left[\right.}+\left.
\frac{M_{\rm ND}}{M_{\rm D}}\frac{\Delta N_{\rm ND}\mp(\Delta N_{\rm ND})^
{1/2}}{N_{\rm ND}\Delta\phi}\right]~~; \\
\slabel{eq:psisd}
&& \sigma_{\Delta{\rm N}/{\rm N}}=\frac{M_{\rm KD}}{M_{\rm D}}
\frac{(\Delta N_{\rm KD})^{1/2}}{N_{\rm KD}}+\frac{M_{\rm ND}}{M_{\rm D}}
\frac{(\Delta N_{\rm ND})^{1/2}}{N_{\rm ND}}
\label{seq:psis}
\end{leftsubeqnarray}
where $M_{\rm KD}$ and $M_{\rm ND}$ are
the SN thick disk and the SN thin disk mass,
$M_{\rm D}=M_{\rm KD}+M_{\rm ND}$, $N_{\rm KD}$
and $N_{\rm ND}$ the number of related sample
objects, and $\Delta N_{\rm KD}$,
$\Delta N_{\rm ND}$, the number of
related sample objects within a
selected metallicity bin, $\Delta
$[O/H], converted into $\Delta\phi$
using Eq.\,(\ref{seq:fib}).   For
further details refer to the parent
papers (C07, C08).

In general, the disk is usually conceived
as made of two main subsystems: the thick
disk and the thin disk.   Accordingly, the
EDOD related to the SN disk depends, via
Eqs.\,(\ref{seq:psis}), on the SN thick to
thin disk mass ratio, $M_{\rm KD}/M_{\rm ND}$,
which is poorly known at present.   Values
already quoted in literature span a wide
range, from some percent (e.g., \cite{hoa07})
to about unity (e.g., \cite{fuh08}),
or even indeterminate in the sense that no
distinction can be made (e.g., \cite{nor87};
\cite{iva08}).   In addition, the global
thick to thin disk mass ratio could
be different from its SN counterpart,
$M_{\rm KD}/M_{\rm ND}$.

\begin{figure*}[t]
\begin{center}
\includegraphics[scale=0.8]{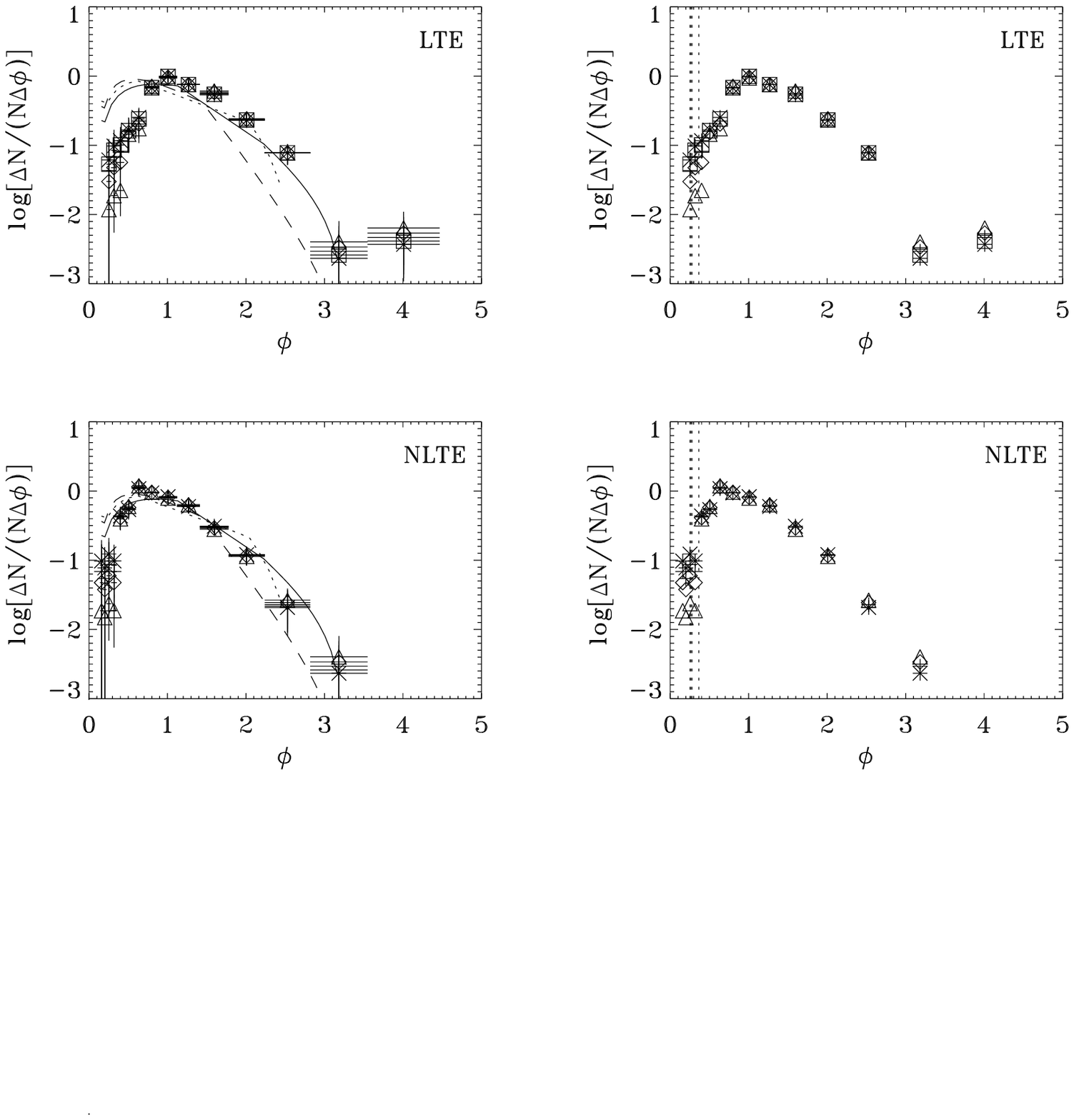}
\caption[DIDOH]{The empirical, differential oxygen abundance
distribution (EDOD) in the solar neighbourhood (SN) thick +
thin disk, deduced from RaK07 and
RaN07 subsamples, both in presence (LTE, top panels)
and in absence (NLTE, bottom panels) of the local thermodynamic
equilibrium approximation.   Different symbols are related to
different thick to thin disk mass ratios, $M_{\rm KD}/M_{\rm ND}$.
Caption of symbols: $M_{\rm KD}/M_{\rm ND}=0.1$ (triangles); 0.3
(diamonds); 0.5 (crosses); 0.7 (squares); 0.9 (asterisks).
Error bars are omitted in right panels to gain clarity.
The theoretical, differential
oxygen abundance distribution (TDOD) related to inhomogeneous
models of chemical evolution for the SN
thick + thin disk (cases KN), is plotted as dashed (case KN2) and full
(case KN3) curves.   Its counterpart, related to the SN thick disk
(cases DB), is plotted as dotted (case DB105) curves.
The dashed vertical bands on right panels correspond to [Fe/H]$=-1$
and related
uncertainties, deduced from Eq.\,(\ref{eq:OFe}) with regard to
a thick disk and a thin disk subsample (\cite{raa07})}
\label{f:DIDOH}
\end{center}
\end{figure*}

The cases, $M_{\rm KD}/M_{\rm ND}=0.1$, 0.3,
0.5, 0.7, 0.9, deduced from RaK07 and RaN07
subsamples, are represented in Fig.\,\ref
{f:DIDOH} as triangles, diamonds, crosses,
squares, asterisks, respectively, both in
presence (LTE) and in absence (NLTE) of the
LTE approximation.
Error bars are plotted in left
panels only, to allow comparison.

The cases, $M_{\rm KD}/M_{\rm ND}=0.1$
and 0.9, deduced from MaK08 and MaN08
subsamples, are represented in Fig.\,\ref
{f:TDBA} (bottom left) as diamonds and
triangles, respectively.

An inspection of Figs.\,\ref{f:TDBA}
(bottom left) and \ref{f:DIDOH} shows
that, in both cases, the uncertainty
on the thick to thin disk mass ratio
is comparable with the error box,
except for the low-metallicity tail
([Fe/H]$<-$1) and/or poorly populated
oxygen abundance bins, where a weak
statistical relevance is expected.

\section{Inhomogeneous, simple models}
\label{s:ism}

In the light of inhomogeneous simple models of chemical
evolution, a selected system is conceived
as being structured into a number of discrete, entirely
gaseous, identical regions, and a background of
long-lived stars, stellar remnants, and gas inhibited from
star formation,
which have been generated earlier. The evolution
occurs via a sequence of identical time steps. At the
beginning of each step, star formation stochastically takes
place in a subclass of ``active'' regions, as described
by simple homogeneous models, while the others remain
``quiescent''. At the end of each step, high-mass stars
have died whereas low-mass stars have survived up until today,
according to instantaneous recycling approximation.
In addition, the enriched gas which remains from active regions
is instantaneously mixed with the unenriched gas
within quiescent regions, to form a new set of identical
regions for the next step.   For further details and
complete formulation refer to earlier attempts (C01,
C07, C08) and parent papers (\cite{maa93}; C00).
Due to reasons of simplicity, the current investigation shall
be restricted to the special case of expected evolution,
where the fraction of active regions is time independent.
For the general case, refer to earlier
attempts (C00, C01).

Input parameters
of inhomogeneous models considered in
the current attempt and parent papers
(C00, C01, C07, C08) are: the number
of steps during and after the assembling
phase; the IMF
power-law exponent; the solar oxygen
abundance; the minimum (beginning of
the first step) and maximum (end of the
last step) normalized oxygen abundance;
the maximum normalized oxygen abundance
at the end of the first step after the
assembling phase; the
true and the effective yield; the
total number of regions at any step
(if taken constant during the evolution)
or the mass of a region (if taken
constant during the evolution); the
total number of steps; as shown in
Table \ref{t:input}.   The more relevant
output parameters are listed in Tables
\ref{t:KN}, \ref{t:DHB}, and \ref{t:IMFU}.
A change in power-law IMF exponent
affects only the mass fraction of
a star generation, $\alpha$, which
remains locked up in long-lived stars
and stellar remnants,
and the lower stellar mass limit,
$m_{\rm mf}$, while the remaining
parameters are left unchanged.   For
further details see the parent papers
(C01, C07, C08).

Values of input parameters for different models related
to the thick disk (models DB), the thin disk (models DN),
the thick + thin disk (models KN), the halo (models H),
and the bulge (models B) are listed in Table \ref{t:input}.
\begin{table*}
\caption[par]{Values of input parameters related to the
expected evolution of inhomogeneous simple models, for
a number of different cases concerning the thick disk
(models DB), the thin disk (models DN), the thick +
thin disk (models KN), the halo (models H), and the bulge
(models B).   The expected evolution is independent of
the total number of regions (C00).   The assumed solar
oxygen abundance is $(Z_{\rm O})_\odot=0.0056$.
The duration of evolution is
needed only for specifying the age-metallicity relation
(C07, C08).   Different values of the IMF power-law
exponent, $p$, leave the output parameters unchanged
with the exception of the mass fraction of a star
generation, $\alpha$, which remains locked up in
long-lived stars and stellar remnants, and the lower stellar
mass limit, $m_{\rm mf}$, as shown in Tables \ref{t:KN},
\ref{t:DHB}, and \ref{t:IMFU}.   Symbol captions: $L_
{\rm C}$ - number of steps during the assembling phase;
$L_{\rm A}$ - number of steps during the whole evolution;
$\phi_{\rm min}$ - minimum
normalized oxygen abundance; $\phi_{\rm max}$ - maximum
normalized oxygen abundance; $\phi_{\rm max}^\ast$ -
maximum normalized oxygen abundance during the $(L_
{\rm C}+1)$-th step; $a$ - slope of the empirical
differential oxygen abundance distribution of the
adjoint simple homogeneous model, see Eq.\,(\ref
{eq:a});
%
%
$\hat{p}$ - true
yield; $\hat{p}^{\prime\prime}$ - effective yield.
%
%
The high number of digits in $\phi_{\rm max}^\ast$ is
necessary to ensure the same IMF during and after the
assembling phase.
%
%
For further details refer to the parent
papers (C00, C01, C07)}
\label{t:input}
\begin{center}
\begin{tabular}{llrlllll}
\multicolumn{1}{c}{case} & \multicolumn{1}{c}{$L_{\rm C}$} &
\multicolumn{1}{c}{$L_{\rm A}$} & \multicolumn{1}{c}{$\phi_{\rm min}$} &
\multicolumn{1}{c}{$\phi_{\rm max}$}  &
\multicolumn{1}{c}{$\phi_{\rm max}^\ast$} &
\multicolumn{1}{c}{$-a\hat{p}^{\prime\prime}/\hat{p}$} &
\multicolumn{1}{c}{$\hat{p}^{\prime\prime}/\hat{p}$} \\
\noalign{\smallskip}
\hline\noalign{\smallskip}
DB12  & 2 & 4  & 0.160 & 2.5   & 2.325653322        & 0.48 & 0.40 \\
DB13  & 2 & 4  & 0.160 & 2.5   & 2.303503529        & 0.52 & 0.40 \\
DB14  & 2 & 4  & 0.160 & 2.5   & 2.284610521        & 0.56 & 0.40 \\
DB15  & 2 & 4  & 0.160 & 2.5   & 2.268320468        & 0.60 & 0.40 \\
DN    & 5 & 25 & 0.400 & 3.2   & 2.01018844         & 0.60 & 0.40 \\
DN4   & 5 & 25 & 0.400 & 3.2   & 2.38931821         & 0.24 & 0.60 \\
DN5   & 5 & 25 & 0.400 & 3.2   & 2.13331046         & 0.30 & 0.60 \\
DN6   & 5 & 25 & 0.400 & 3.2   & 1.96266176         & 0.36 & 0.60 \\
DN7   & 5 & 25 & 0.400 & 3.2   & 1.84079119         & 0.42 & 0.60 \\
KN1   & 5 & 25 & 0.160 & 3.2   & 2.09520539         & 0.46 & 0.40 \\
KN2   & 5 & 25 & 0.160 & 3.2   & 1.83493684         & 0.60 & 0.40 \\
KN3   & 5 & 25 & 0.160 & 3.2   & 1.59805348         & 0.46 & 0.60 \\
KN4   & 5 & 25 & 0.160 & 2.5   & 1.90667713         & 0.46 & 0.40 \\
KN5   & 5 & 25 & 0.160 & 2.5   & 1.420631883        & 0.46 & 0.60 \\
B1    & 2 & 4  & 0.200 & 5.5   & 4.44130682         & 0.40 & 0.70 \\
H1    & 2 & 4  & 0.001 & 1.0   & 0.9834230800463403 & 3.25 & 0.04 \\
DB105 & 2 & 4  & 0.160 & 2.5   & 2.367027589        & 0.42 & 0.40 \\
DN81  & 5 & 25 & 0.200 & 2.5   & 1.51408616         & 0.42 & 0.60 \\
KN6   & 5 & 25 & 0.160 & 2.5   & 2.01304781         & 0.42 & 0.40 \\
B     & 2 & 4  & 0.200 & 5.5   & 4.43357173         & 0.42 & 0.70 \\
H     & 2 & 4  & 0.100 & 1.795 & 1.7907654704       & 0.42 & 0.40 \\
\noalign{\smallskip}
\hline
\end{tabular}
\end{center}
\end{table*}
Models B1 and H1 come from earlier work (C07).

Available samples, from which the EDOD has been
deduced, are biased (towards low metallicities)
and/or incomplete: for this reason, little meaning
would be related to the best fit to the data.
Different models considered here are mainly aimed
in (i) analysing the effect of different input
parameters on oxygen enrichment; (ii) providing
acceptable fits to the data; (iii) investigating
if a class of models with strictly universal IMF exist,
which reproduce the EDOD determined for different
populations.

An equal number of steps has been chosen for both
the assembling phase and the whole evolution, with
regard to the halo, the bulge, and the thick disk,
which formed within about 1 Gyr (e.g., \cite{pra08};
\cite{zoa08}; \cite{mea08}).   On the other hand,
a larger number of steps has been assumed for the
thin disk, which is still undergoing chemical
evolution.   Accordingly, 4 episodes of star
formation (2 during the assembling phase) have
been (somewhat arbitrarily) chosen for the halo,
the bulge, and the thick disk, and 25 (5 during
the assembling phase) for the thin disk.

The initial and final normalized oxygen abundance
have been chosen equal to the minimum and maximum
value related to the most populated sample.   With
regard to the thin and the thick + thin disk, the
above mentioned values are changed in some models
(DN81, KN4, KN5, KN6) for improving the fit to the
data.   The maximum normalized oxygen abundance
during the ($L_C$+1)-th step, $\phi_{\rm max}^\ast$,
is fixed by the boundary condition of equal IMF
during and after the assembling phase.
The effective yield to the true yield ratio,
$\hat{p}^{\prime\prime}/\hat{p}$, and the slope
of the TDOD of the adjoint simple homogeneous
model, $a$, have been selected within a narrow
range, outside which no acceptable fit to the
data has been found.
More specifically, the following considerations
can be made.

The lower and upper normalized oxygen abundance
limit, $\phi_{\rm min}$ and $\phi_{\rm max}$,
have to be conceived as boundary conditions
with respect to the subsystem under consideration.
Assuming, say, $\phi_{\rm min}=0$, would be
meaningless in that (i) no star with zero
metallicity has still been detected, and (ii)
it would imply a G-dwarf problem in absence
of ad hoc assumptions.   Values $(\phi_{\rm min},
\phi_{\rm max})=(0.16,2.5)$ for the SN thick
disk, (0.4,\,3.2) for the SN thin disk, (0.16,\,3.2)
for the SN thick + thin disk, (0.001,\,1.0) for
the SN halo, (0.2,\,5.5) for the bulge are taken
from the most populated sample from which the
EDOD has been deduced.   Values (0.2,\,2.5) for
the SN thin disk and (0.16,\,2.5) for the SN
thick + thin disk are obtained by removing a
single oxygen overabundant star (\astrobj{HIP68184})
from the RaN07 subsample and, in the former
case, assuming the low-metallicity tail of
SN (still undetected) thin disk stars extends
down to $\phi=0.2$.   Values (0.1,\,1.795)
for the SN inner halo are taken from the
incomplete RaH07 subsample, which makes the
associated model poorly representative, even
if conceptually interesting.

The maximum normalized oxygen abundance
during the $(L_{\rm C}+1)$-th step,
$\phi_{\rm max}^\ast$, defines the IMF
during (first $L_{\rm C}$ steps) and
after (subsequent $L_{\rm A}-L_{\rm C}$ steps)
the assembling phase.   The high number
of digits ensures the same IMF during
the whole evolution.   It can be seen that
the same number of steps during and after
the assembling phase, has been assigned to
both the SN thick disk, the SN halo, and
the bulge, on one hand, and  to both the
SN thin and the SN thick + thin disk, on
the other hand.   Strictly speaking,
it would not imply a similar formation
timescale, as suggested by recent findings
(e.g., \cite{mea08}): to this
aim, an equal duration would also be
assigned to each step, which is relevant
in dealing with the age-metallicity
relation, but leaves the TDOD unchanged.

The remaining input parameters are the
effective to true yield ratio, $\hat{p}^
{\prime\prime}/\hat{p}$, and the product:
\begin{equation}
\label{eq:a}
-a\frac{\hat{p}^{\prime\prime}}{\hat{p}}=\frac{(Z_{\rm O})_\odot}
{\ln10}\frac1{\hat{p}}~~;
\end{equation}
which is inversely proportional to the true
yield.   For further details
refer to parent papers (C00, C01, C07).

Values of output parameters for models shown
in Table \ref{t:input} are listed in Tables
\ref{t:KN} and \ref{t:DHB}, respectively.
Additional cases where gas in neither inhibited
from, nor enhanced in, star formation, are
listed in Table \ref{t:IMFU}.
\begin{table*}
%
%
\caption[par]
{Values of output parameters related to the
expected evolution of inhomogeneous simple models, for
four different cases concerning the thick disk (models
DB), and five concerning the thin disk (models DN),
respectively.   The indices, 2.9 and 2.35,
denote values related to the power-law IMF
exponent, $p$, in computing the corresponding
quantities.   For the parameter,
$\psi_1$, upper and lower values are
calculated as in an earlier attempt (C07)
by use of Eqs.\,(40)
and (41) therein, respectively.   The
effective yield, $\hat{p}^\prime$, is
related to inhomogenities in oxygen
abundance due to the presence of active
and quiescent regions, whereas oxygen is
uniformly distributed within active regions.
The lower part of the Table (last 6 rows)
is related to models with inhibited
star formation.   Parameters not reported
therein have the same value as in the upper part,
with the exception of $\hat{p}$, $\alpha$,
and $\widetilde{m}_{\rm mf}=m_{\rm mf}/{\rm m}_\odot$,
which are listed in Table \ref
{t:IMFU} together with other parameters not
appearing here.   The effective yield, $\hat
{p}^{\prime\prime}$, due to the presence of
(inhibiting star formation) gas within active
regions, is listed as $\hat{p}$ in the upper 
part of the table.
The effective yield, $\hat{p}^\prime$, due to
the presence of both (inhibiting
star formation) gas within active
regions, and (precluding star formation) gas
within quiescent regions, is listed with the
same notation in the upper part of the table.
The index, R, denotes a generic active
region.   The mean oxygen abundance
(normalized to the solar value) of stars
at the end of evolution is denoted as $\overline
{\phi}$}
\label{t:KN}
%
%
\begin{center}
%
%
\begin{tabular}{llllllrrrl}
\multicolumn{1}{c|}{\phantom{$\phi$}}
&\multicolumn{4}{c|}{LTE}
&\multicolumn{4}{c|}{NLTE} \\
\hline\noalign{\smallskip}
\multicolumn{1}{c}{} &
\multicolumn{1}{c}{DB12} & \multicolumn{1}{c}{DB13} &
\multicolumn{1}{c}{DB14} & \multicolumn{1}{c}{DB15} &
\multicolumn{1}{c}{DN}  &
\multicolumn{1}{c}{DN4} &
\multicolumn{1}{c}{DN5} &
\multicolumn{1}{c}{DN6} & \multicolumn{1}{c}{DN7}  \\
\noalign{\smallskip}
\hline\noalign{\smallskip}
$\mu_{\rm R}^\prime$               & 1.3243~E$-$1 & 1.2295~E$-$1 & 1.1258~E$-$1 & 1.0304~E$-$1 & 1.6663~E$-$1 &    3.7477~E$-$1 &    3.6662~E$-$1 &    3.5864~E$-$1 & 3.5083~E$-$1 \\
$q$                                & 6.1771~E$-$1 & 5.5533~E$-$1 & 4.9941~E$-$1 & 4.4924~E$-$1 & 8.0550~E$-$1 &    9.6146~E$-$1 &    9.3741~E$-$1 &    9.1396~E$-$1 & 8.9109~E$-$1 \\
$\chi$                             & 4.4156~E$-$1 & 5.0700~E$-$1 & 5.6410~E$-$1 & 6.1403~E$-$1 & 2.3339~E$-$1 &    6.1635~E$-$2 &    9.8821~E$-$2 &    1.3416~E$-$1 & 1.6776~E$-$1 \\
$\hat{p}/(Z_{\rm O})_\odot$        & 9.0478~E$-$1 & 8.3518~E$-$1 & 7.7553~E$-$1 & 7.2382~E$-$1 & 7.2382~E$-$1 &    1.8096~E$-$0 &    1.4476~E$-$0 &    1.2064~E$-$0 & 1.0340~E$-$0 \\
$\hat{p}^\prime/(Z_{\rm O})_\odot$ & 3.6191~E$-$1 & 3.3407~E$-$1 & 3.1021~E$-$1 & 2.8953~E$-$1 & 2.8953~E$-$1 &    1.0857~E$-$0 &    8.6859~E$-$1 &    7.2382~E$-$1 & 6.2042~E$-$1 \\
$\mu_f$                            & 1.4559~E$-$1 & 9.5108~E$-$2 & 6.2204~E$-$2 & 4.0731~E$-$2 & 4.4842~E$-$3 &    3.7439~E$-$1 &    1.9872~E$-$1 &    1.0548~E$-$1 & 5.5988~E$-$2 \\
$\alpha_{2.9}$                     & 6.9505~E$-$1 & 7.1175~E$-$1 & 7.2671~E$-$1 & 7.4020~E$-$1 & 7.4020~E$-$1 &    5.3263~E$-$1 &    5.8755~E$-$1 &    6.3092~E$-$1 & 6.6604~E$-$1 \\
$\alpha_{2.35}$                    & 8.6951~E$-$1 & 8.7833~E$-$1 & 8.8803~E$-$1 & 8.9281~E$-$1 & 8.9281~E$-$1 &    7.6914~E$-$1 &    8.0638~E$-$1 &    8.3327~E$-$1 & 8.5360~E$-$1 \\
$(\widetilde{m}_{\rm mf})_{2.9}$       & 4.0227~E$-$1 & 3.7813~E$-$1 & 3.5660~E$-$1 & 3.3728~E$-$1 & 3.3728~E$-$1 &    6.4228~E$-$1 &    5.6022~E$-$1 &    4.9602~E$-$1 & 4.4449~E$-$1 \\
$(\widetilde{m}_{\rm mf})_{2.35}$      & 1.1305~E$-$2 & 9.3463~E$-$3 & 7.8194~E$-$3 & 6.6108~E$-$3 & 6.6108~E$-$3 &    5.1835~E$-$2 &    3.2618~E$-$2 &    2.1898~E$-$2 & 1.5436~E$-$2 \\
$\Delta^\ast\phi$                  & 1.7435~E$-$1 & 1.9650~E$-$1 & 2.1539~E$-$1 & 2.3168~E$-$1 & 6.2622~E$-$2 &    4.2667~E$-$2 &    5.6142~E$-$2 &    6.5123~E$-$2 & 7.1537~E$-$2 \\
$\Delta^\ast\phi_{\rm R}^\prime$   & 1.8170~E$-$0 & 1.7505~E$-$0 & 1.6938~E$-$0 & 1.6450~E$-$0 & 1.2971~E$-$0 &    1.7760~E$-$0 &    1.4526~E$-$0 &    1.2370~E$-$0 & 1.0831~E$-$0 \\
$-\psi_o$                          & 2.4322~E$-$1 & 1.7337~E$-$1 & 1.1035~E$-$1 & 5.3384~E$-$2 & 4.8961~E$-$2 &    1.2641~E$-$0 &    1.0696~E$-$0 &    9.0546~E$-$1 & 7.6482~E$-$1 \\
$-\psi_1(40)$                      & 2.8439~E$-$1 & 2.2346~E$-$1 & 1.6926~E$-$1 & 1.2104~E$-$1 & 5.0826~E$-$1 &    1.2692~E$-$0 &    1.0780~E$-$0 &    9.1713~E$-$1 & 7.7976~E$-$1 \\
$-\psi_1(41)$                      & 2.8506~E$-$1 & 2.2446~E$-$1 & 1.7066~E$-$1 & 1.2289~E$-$1 & 5.0839~E$-$1 &    1.2692~E$-$0 &    1.0780~E$-$0 &    9.1718~E$-$1 & 7.7985~E$-$1 \\
                                   &              &              &              &              &              &                 &                 &                 &              \\
$\kappa$                           & 1.4286~E$-$1 & 2.3810~E$-$1 & 3.3333~E$-$1 & 4.2857~E$-$1 & 4.2857~E$-$1 & $-$4.2857~E$-$1 & $-$2.8571~E$-$1 & $-$1.4286~E$-$1 & 0.0000~E$-$0 \\
$s_{\rm R}^\prime$                 & 7.5755~E$-$1 & 7.0839~E$-$1 & 6.6557~E$-$1 & 6.2787~E$-$1 & 5.8336~E$-$1 &    1.0942~E$-$0 &    8.8673~E$-$1 &    7.4825~E$-$1 & 6.4917~E$-$1 \\
$D_{\rm R}^\prime$                 & 1.0822~E$-$1 & 1.6866~E$-$1 & 2.2186~E$-$1 & 2.6909~E$-$1 & 2.5001~E$-$1 & $-$4.6892~E$-$1 & $-$2.5335~E$-$1 & $-$1.0689~E$-$1 & 0.0000~E$-$0 \\
$\overline{\phi}$                  & 7.8307~E$-$1 & 7.4978~E$-$1 & 7.2065~E$-$1 & 6.9485~E$-$1 & 8.6448~E$-$1 &    1.1450~E$-$0 &    1.0068~E$-$0 &    9.1462~E$-$1 & 8.4870~E$-$1 \\
$s_f$                              & 7.4761~E$-$1 & 7.3087~E$-$1 & 7.0335~E$-$1 & 6.7149~E$-$1 & 6.9686~E$-$1 &    1.0948~E$-$0 &    1.1218~E$-$0 &    1.0436~E$-$0 & 9.4401~E$-$1 \\
$D_f$                              & 1.0680~E$-$1 & 1.7402~E$-$1 & 2.3445~E$-$1 & 2.8778~E$-$1 & 2.9865~E$-$1 & $-$4.6921~E$-$1 & $-$3.2051~E$-$1 & $-$1.4909~E$-$1 & 0.0000~E$-$0 \\
%
%
\noalign{\smallskip}
\hline
\end{tabular}
\end{center}
\end{table*}
%
%
%
\begin{table*}
\caption[par]{Values of output parameters related to the
expected evolution of inhomogeneous simple models, for
five different cases concerning the thick + thin disk
(models KN).   Additional cases related to the bulge
(model B1) and the halo (model H1), taken from an
earlier attempt (C07), but
related to a different reference model with respect to
inhibition from and enhancement in star formation, are
listed for comparison.   Captions as in Table \ref{t:KN}}
\label{t:DHB}
\begin{center}
\begin{tabular}{llllllrl}
\hline\noalign{\smallskip}
\multicolumn{1}{c}{} &
\multicolumn{1}{c}{KN1} & \multicolumn{1}{c}{KN2} &
\multicolumn{1}{c}{KN3} & \multicolumn{1}{c}{KN4} &
\multicolumn{1}{c}{KN5}  &
\multicolumn{1}{c}{B1} &
\multicolumn{1}{c}{H1} \\
$\mu_{\rm R}^\prime$               & 1.7520~E$-$1 & 1.6240~E$-$1 & 3.4073~E$-$1 & 1.8550~E$-$1 & 3.5544~E$-$1 &    1.4139~E$-$1 &    8.2201~E$-$4  \\
$q$                                & 8.5730~E$-$1 & 7.8025~E$-$1 & 8.6171~E$-$1 & 9.2064~E$-$1 & 9.0458~E$-$1 &    2.4833~E$-$1 &    4.4991~E$-$2  \\
$\chi$                             & 1.7302~E$-$1 & 2.6236~E$-$1 & 2.0977~E$-$1 & 9.7438~E$-$2 & 1.4804~E$-$1 &    8.7545~E$-$1 &    9.5579~E$-$1  \\
$\hat{p}/(Z_{\rm O})_\odot$        & 9.4412~E$-$1 & 7.2382~E$-$1 & 9.4412~E$-$1 & 9.4412~E$-$1 & 9.4412~E$-$1 &    1.0857~E$-$0 &    1.3363~E$-$1  \\
$\hat{p}^\prime/(Z_{\rm O})_\odot$ & 3.7765~E$-$1 & 2.8953~E$-$1 & 5.6647~E$-$1 & 3.7765~E$-$1 & 5.6647~E$-$1 &    7.6001~E$-$1 &    5.3452~E$-$3  \\
$\mu_f$                            & 2.1295~E$-$2 & 2.0221~E$-$3 & 2.4210~E$-$2 & 1.2653~E$-$1 & 8.1501~E$-$2 &    3.8030~E$-$3 &    4.0973~E$-$6  \\
$\alpha_{2.9}$                     & 6.8596~E$-$1 & 7.4020~E$-$1 & 6.8596~E$-$1 & 6.8596~E$-$1 & 6.8596~E$-$1 &    6.5510~E$-$1 &    9.3915~E$-$1  \\
$\alpha_{2.35}$                    & 8.6460~E$-$1 & 8.9281~E$-$1 & 8.6460~E$-$1 & 8.6460~E$-$1 & 8.6460~E$-$1 &    8.4739~E$-$1 &    9.7832~E$-$1  \\
$(\widetilde{m}_{\rm mf})_{2.9}$       & 4.1547~E$-$1 & 3.3728~E$-$1 & 4.1547~E$-$1 & 4.1547~E$-$1 & 4.1547~E$-$1 &    4.6050~E$-$1 &    6.7781~E$-$2  \\
$(\widetilde{m}_{\rm mf})_{2.35}$      & 1.2495~E$-$2 & 6.6108~E$-$3 & 1.2495~E$-$2 & 1.2495~E$-$2 & 1.2495~E$-$2 &    1.7263~E$-$2 &    7.5651~E$-$5  \\
$\Delta^\ast\phi$                  & 5.8147~E$-$2 & 7.1945~E$-$2 & 8.4313~E$-$2 & 3.1228~E$-$2 & 5.6809~E$-$2 &    1.0587~E$-$0 &    1.6577~E$-$2  \\
$\Delta^\ast\phi_{\rm R}^\prime$   & 1.6445~E$-$0 & 1.3157~E$-$0 & 1.0165~E$-$0 & 1.5905~E$-$0 & 9.7659~E$-$1 &    2.1239~E$-$0 &    9.4927~E$-$1  \\
$-\psi_o$                          & 7.2759~E$-$1 & 4.3986~E$-$1 & 6.4265~E$-$1 & 9.2754~E$-$1 & 7.6772~E$-$1 &    9.1837~E$-$2 &    8.5447~E$-$1  \\
$-\psi_1(40)$                      & 7.4089~E$-$1 & 4.6123~E$-$1 & 6.6189~E$-$1 & 9.3471~E$-$1 & 7.8072~E$-$1 &    2.8650~E$-$1 &    8.2781~E$-$1  \\
$-\psi_1(41)$                      & 7.4096~E$-$1 & 4.6141~E$-$1 & 6.6204~E$-$1 & 9.3473~E$-$1 & 7.8079~E$-$1 &    3.0358~E$-$1 &    8.2753~E$-$1  \\
                                   &              &              &              &              &              &                 &                  \\
$\kappa$                           & 9.5238~E$-$2 & 4.2857~E$-$1 & 9.5238~E$-$2 & 9.5238~E$-$2 & 9.5238~E$-$2 & $-$4.7619~E$-$2 &    6.7381~E$-$0  \\
$s_{\rm R}^\prime$                 & 7.5307~E$-$1 & 5.8632~E$-$1 & 6.0194~E$-$1 & 7.4367~E$-$1 & 5.8851~E$-$1 &    9.0154~E$-$1 &    1.2912~E$-$1  \\
$D_{\rm R}^\prime$                 & 7.1721~E$-$2 & 2.5128~E$-$1 & 5.7327~E$-$2 & 7.0826~E$-$2 & 5.6048~E$-$2 & $-$4.2930~E$-$2 &    8.7005~E$-$1  \\
$\overline{\phi}$                  & 7.5480~E$-$1 & 6.2873~E$-$1 & 5.7876~E$-$1 & 7.4187~E$-$1 & 5.6558~E$-$1 &    9.3597~E$-$1 &    1.3385~E$-$1  \\
$s_f$                              & 8.9360~E$-$1 & 6.9858~E$-$1 & 8.9094~E$-$1 & 7.9751~E$-$1 & 8.3863~E$-$1 &    1.0460~E$-$0 &    1.2923~E$-$1  \\
$D_f$                              & 8.5105~E$-$2 & 2.9939~E$-$1 & 8.4851~E$-$2 & 7.5954~E$-$2 & 7.9869~E$-$2 & $-$4.9810~E$-$2 &    8.7077~E$-$1  \\
\noalign{\smallskip}
\hline
\end{tabular}
\end{center}
\end{table*}
\begin{table*}
\caption[par]{Values of output parameters related to the
expected evolution of inhomogeneous simple models, for
five different cases concerning the thick disk (model
DB105), the thin disk (model DN81), the thick + thin
disk (model KN6), the bulge (model B), and the inner halo
(model H).   The stellar initial
mass function (IMF) is strictly universal, in the sense
that neither inhibition from, nor enhancement in, star
formation, appear in one model with respect to the
other.   Captions as in Table \ref{t:KN}}
\label{t:IMFU}
\begin{center}
\begin{tabular}{llllllll}
\hline\noalign{\smallskip}
\multicolumn{1}{c}{} &
\multicolumn{1}{c}{DB105} & \multicolumn{1}{c}{DN81} &
\multicolumn{1}{c}{KN6} & \multicolumn{1}{c}{B} &
\multicolumn{1}{c}{H}  \\
$\mu_{\rm R}^\prime$               & 1.5302~E$-$1 & 3.6062~E$-$1 & 1.8860~E$-$1 & 1.3113~E$-$1 & 1.9653~E$-$1  \\
$q$                                & 7.2507~E$-$1 & 9.1977~E$-$1 & 9.3992~E$-$1 & 2.2916~E$-$1 & 9.8981~E$-$1  \\
$\chi$                             & 3.2460~E$-$1 & 1.2549~E$-$1 & 7.4049~E$-$2 & 8.8717~E$-$1 & 1.2677~E$-$2  \\
$\hat{p}/(Z_{\rm O})_\odot$        & 1.0340~E$-$0 & 1.0340~E$-$0 & 1.0340~E$-$0 & 1.0340~E$-$0 & 1.0340~E$-$0  \\
$\hat{p}^\prime/(Z_{\rm O})_\odot$ & 4.1361~E$-$1 & 6.2042~E$-$1 & 4.1361~E$-$1 & 7.2382~E$-$1 & 4.1361~E$-$1  \\
$\mu_f$                            & 2.7639~E$-$1 & 1.2357~E$-$1 & 2.1244~E$-$1 & 2.7579~E$-$3 & 9.5988~E$-$1  \\
$\alpha_{2.9}$                     & 6.6604~E$-$1 & 6.6604~E$-$1 & 6.6604~E$-$1 & 6.6604~E$-$1 & 6.6604~E$-$1  \\
$\alpha_{2.35}$                    & 8.5360~E$-$1 & 8.5360~E$-$1 & 8.5360~E$-$1 & 8.5360~E$-$1 & 8.5360~E$-$1  \\
$(\widetilde{m}_{\rm mf})_{2.9}$       & 4.4449~E$-$1 & 4.4449~E$-$1 & 4.4449~E$-$1 & 4.4449~E$-$1 & 4.4449~E$-$1  \\
$(\widetilde{m}_{\rm mf})_{2.35}$      & 1.5436~E$-$2 & 1.5436~E$-$2 & 1.5436~E$-$2 & 1.5436~E$-$2 & 1.5436~E$-$2  \\
$\Delta^\ast\phi$                  & 1.3297~E$-$1 & 5.1890~E$-$2 & 2.5629~E$-$2 & 1.0664~E$-$0 & 4.2345~E$-$3  \\
$\Delta^\ast\phi_{\rm R}^\prime$   & 1.9411~E$-$0 & 1.0546~E$-$0 & 1.7249~E$-$0 & 2.1007~E$-$0 & 1.6823~E$-$0  \\
$-\psi_o$                          & 3.6269~E$-$1 & 8.5864~E$-$1 & 1.0413~E$-$0 & 6.5328~E$-$2 & 5.1492~E$-$1  \\
$-\psi_1(40)$                      & 3.9032~E$-$1 & 8.6950~E$-$1 & 1.0467~E$-$0 & 2.7020~E$-$1 & 5.1581~E$-$1  \\
$-\psi_1(41)$                      & 3.9062~E$-$1 & 8.6954~E$-$1 & 1.0467~E$-$0 & 2.8928~E$-$1 & 5.1581~E$-$1  \\
                                   &              &              &              &              &               \\
$\kappa$                           & 0.0000~E$-$0 & 0.0000~E$-$0 & 0.0000~E$-$0 & 0.0000~E$-$0 & 0.0000~E$-$0  \\
$s_{\rm R}^\prime$                 & 8.4698~E$-$1 & 6.3938~E$-$1 & 8.1140~E$-$1 & 8.6887~E$-$1 & 8.0347~E$-$1  \\
$D_{\rm R}^\prime$                 & 0.0000~E$-$0 & 0.0000~E$-$0 & 0.0000~E$-$0 & 0.0000~E$-$0 & 0.0000~E$-$0  \\
$\overline{\phi}$                  & 8.4335~E$-$1 & 6.3920~E$-$1 & 7.9310~E$-$1 & 9.1700~E$-$1 & 7.2253~E$-$1  \\
$s_f$                              & 7.2361~E$-$1 & 8.7643~E$-$1 & 7.8756~E$-$1 & 9.9724~E$-$1 & 4.0244~E$-$2  \\
$D_f$                              & 0.0000~E$-$0 & 0.0000~E$-$0 & 0.0000~E$-$0 & 0.0000~E$-$0 & 0.0000~E$-$0  \\
\noalign{\smallskip}
\hline
\end{tabular}
\end{center}
\end{table*}
The meaning of the parameters is explained in 
Appendix \ref{a:simbo}.   For further details refer
to parent papers (C00, C01, C07).

The TDOD related to models DB15 and DN is compared
in Fig.\,\ref{f:OKND} to the corresponding EDOD
with regard to SN thick (RaK07 subsample, left panels)
and SN thin (RaN07 subsample, right panels) disk
stars, respectively, both in presence (top panels)
and in absence (bottom panels) of the LTE approximation.
On each panel, the fit to the data is plotted
as a full curve, and its counterpart related to the
right or left panel as a dashed line for comparison.
Models are fitted to NLTE data.

The TDOD related to models DB15, DN81, KN4, and B,
is compared in Fig.\,\ref{f:TDBA} to the corresponding
EDOD with regard to SN thick disk (MaK08 subsample,
top left), SN thin disk (MaN08 subsample, top right),
SN thick + thin disk [MaK08 + MaN08 subsample with
assumed $M_{\rm KD}/M_{\rm ND}=0.1$ (diamonds) and
0.9 (triangles), bottom left], and bulge (MaB08
subsample, bottom right) stars, respectively.
On each panel, the fit to the data is plotted
as a full curve, and its counterpart related to the
right or left panel as a dotted line for comparison.
Model B1 is represented by the dashed curve (bottom
right).

The TDOD related to models DB15 (full), DN
(dot-dashed), B1 (dotted), and H1 (dashed),
is compared in Fig.\,\ref{f:THUD} to the
corresponding EDOD with regard to SN
uncertain population (RaU07 subsample,
left panels) and SN halo (RaH07
subsample, right panels) stars,
respectively, both in presence (top
panels) and in absence (bottom panels)
of the LTE approximation.

The TDOD related to models KN2 (dashed)
and KN3 (full) is compared in Fig.\,\ref
{f:DIDOH} to the corresponding EDOD,
with regard to SN thick + thin disk
(RaK07 + RaN07 subsample) stars for
assumed SN thick to thin disk mass
ratio, $M_{\rm KD}/M_{\rm ND}=0.1$
(triangles), 0.3 (diamonds), 0.5
(crosses), 0.7 (squares), and 0.9
(asterisks).   Both error bars and
TDODs are omitted in right panels
to gain clarity.   Model DB105, related
to the SN thick disk, is also represented
as a dotted line.   The dashed vertical
bands correspond to [Fe/H]$=-$1 and
related uncertainties, according to
Eq.\,(\ref{eq:OFe}), with regard to
a thick disk (left) and a thin disk
(right) subsample (\cite{raa07}).
Models are fitted to NLTE
data.

\begin{figure*}[t]
\begin{center}
\includegraphics[scale=0.8]{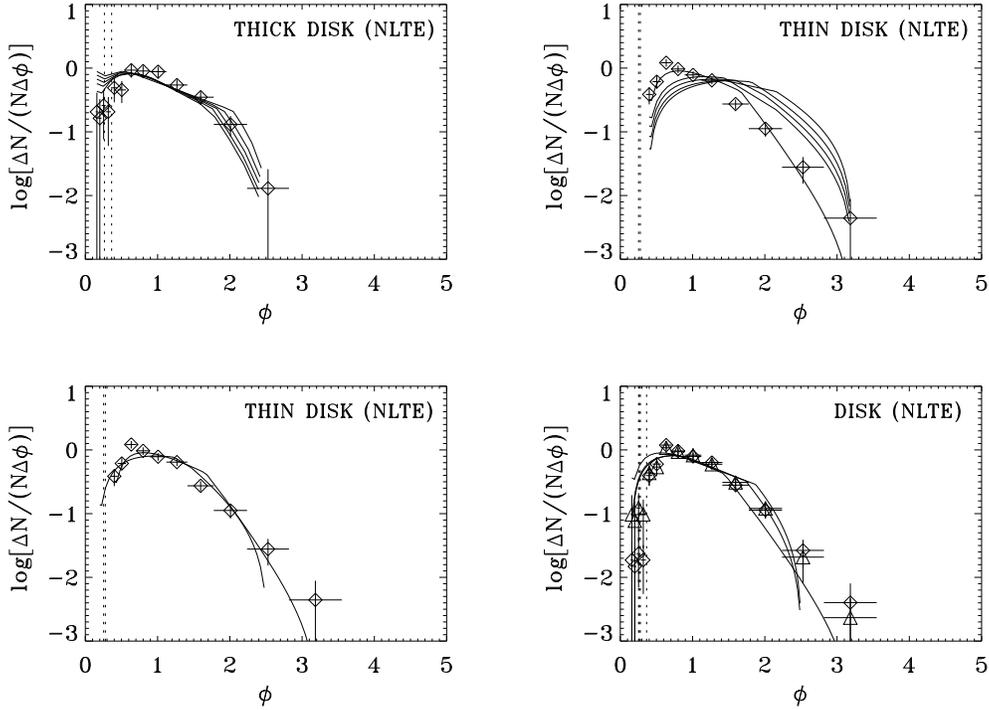}
\caption[OKNDA]{The empirical, differential oxygen abundance
distribution (EDOD) related to the RaK07 (top left),
RaN07 (top right and bottom left) and RaK07 + RaN07
(bottom right) subsample, for assumed SN thick to
thin disk mass ratio, $M_{\rm KD}/M_{\rm ND}=0.1$
(diamonds) and 0.9 (triangles), in absence of the
local thermodynamic equilibrium (LTE) approximation.
The theoretical, differential
oxygen abundance distribution (TDOD) related to inhomogeneous
models of chemical evolution, is plotted for the following
models: DB12, DB13, DB14, DB15, DB105 (from up to down
along the vertical line, $\phi=2.0$, top left);
DN4, DN5, DN6, DN7, DN (from up to
down along the vertical line, $\phi=
2.5$, top right); DN, DN81 (from up to down
along the vertical line, $\phi=2.4$, bottom left);
KN2, KN4, KN6 (from down to up
along the vertical line, $\phi=2.0$, bottom right).
The dashed vertical bands correspond to [Fe/H]$=-1$ and related
uncertainties, deduced from Eq.\,(\ref{eq:OFe}) with regard to
a thick disk (upper left and lower right) and a thin disk
(upper right and bottom panels) subsample (\cite{raa07})}
\label{f:OKNDA}
\end{center}
\end{figure*}

The TDOD related to models DB12,
DB13, DB14, DB15, DB105 (from up to down
along the vertical line, $\phi=2.0$, top left);
DN4, DN5, DN6, DN7, DN (from up to
down along the vertical line, $\phi=
2.5$, top right); DN, DN81 (from up to down
along the vertical line, $\phi=2.4$, bottom left);
KN2, KN4, KN6 (from down to up
along the vertical line, $\phi=2.0$, bottom right);
is compared in Fig.\,\ref{f:OKNDA} to the
corresponding EDOD with regard to SN
thick disk (RaK07 subsample, top left),
SN thin disk (RaN07 subsample, top right
and bottom left), and SN thick + thin
disk [RaK07 + RaN07 subsample with
assumed $M_{\rm KD}/M_{\rm ND}=0.1$ (diamonds) and
0.9 (triangles), bottom right]
stars, respectively,
in absence of the LTE approximation.

\section{Discussion} \label{s:disc}
\subsection{Fitting empirical differential oxygen
abundance distributions} \label{ss:ETD}

Acceptable fits to the EDOD related to
SN thin or thick + thin disk stars,
must satisfy the boundary condition
of a nonnegligible present-day gas
mass fraction, $\mu_f=M_{\rm g}/M=
0.1$-0.3 (e.g.,  \cite{pra95}).
Accordingly, related models
(DN and KN, respectively) must
be disregarded if (i) the EDOD
is poorly fitted and/or (ii)
$\mu_f$ lies outside the above
mentioned range.

With regard to the SN thin disk,
an inspection of Figs.\,\ref{f:OKND},
\ref{f:THUD}, \ref{f:OKNDA} and
Table \ref{t:KN} shows that the
sole acceptable model (DN81) needs
lowered initial and final oxygen
abundance, close to thick disk values.
More specifically, fitting the data
implies the existence of SN thin
disk stars with normalized oxygen
abundance down to $\phi\approx0.2$
and, in addition, the removal of a
single oxygen overabundant 
([O/H]$_{\rm NLTE}=0.48$ or $\phi
\approx3$) star (\astrobj{HIP68184})
from the sample, which could be due
to peculiar evolutionary effects
and/or systematic errors.
In the light of the model, the
following conclusion holds: if
the thick and the thin disk
evolved as separate entities,
the chemical evolution spanned
along a similar oxygen abundance
range.

With regard to the SN thick +
thin disk, an inspection of
Fig.\,\ref{f:OKNDA} and Table
\ref{t:DHB} shows that the two
acceptable models (KN4 and KN6)
need lowered final oxygen abundance,
close to thick disk values.
More specifically, fitting the
data implies the removal of the
above mentioned, oxygen overabundant
star, from the sample.   In the
light of the model, the following
conclusion holds: if the thick
and the thin disk evolved as a single
entity, the chemical evolution
spanned along a similar oxygen
abundance range with respect to
the thick disk considered alone.
Models where the thick and the
thin disk evolve as a whole,
are consistent with the existence
of an evolutionary link between
the two disk components (\cite{hay08}).
It is worth remembering that the
above quoted paper is referring
to a kinematic evolution or
migration between thin and thick disks,
as opposed to some form of Galactic
chemical evolution.

With regard to the bulge, an inspection
of Figs.\,\ref{f:TDBA}, \ref{f:THUD}
and Tables \ref{t:DHB}, \ref{t:IMFU}
shows that both the models considered
(B and B1) provide acceptable fits to
bulge and SN halo data, with negligible
amount of gas left at the end of
evolution, as expected.   In the light
of the model, the following conclusion
holds: oxygen production within the
bulge and the SN halo underwent a
similar history for the common abundance
range.   This result is highly speculative,
due to the incompleteness of the halo
sample from which the data were extracted
(\cite{raa07}).

With regard to the SN halo, an inspection
of Fig.\,\ref{f:THUD} and Tables \ref{t:DHB},
\ref{t:IMFU}  shows that a fraction of
low-metallicity data is fitted by a single
model (H1), while the remaining part is
fitted by a second model (H).   Accordingly,
the (incomplete) halo sample from which the
data were extracted (\cite{raa07}) has to
be conceived as made of two distinct classes
of objects, namely (i) stars from a gas
reservoir from which both the inner halo,
the bulge, and the thick disk originated, and
(ii) stars from a gas reservoir from which
the outer halo originated.   In the light
of the model, the following conclusion holds:
oxygen production within the SN inner halo,
the bulge, and the thick disk underwent a
similar history for the common abundance
ranges, while the chemical evolution of the
SN outer halo was different.   In fact,
input parameters related to the outer halo
model (H1) are markedly different from the
remaining ones, as shown in Table \ref{t:input}.
A stellar dichotomy between the inner and the
outer halo has recently been detected (\cite{caa07}).

The assumption of universal IMF implies
gas inhibition from, or enhancement in,
star formation, with respect to a
reference thin disk model (C07, C08),
which has been chosen to be DN81.
Then a legitimate question
is about the existence of models
which fit to an acceptable extent
the EDOD related to the SN thick
disk, SN thin disk, SN inner halo,
bulge, and
where, in addition, gas is neither
inhibited from, nor enhanced in,
star formation.   The answer is
positive, as shown in Table
\ref{t:IMFU} and Figs.\,\ref{f:TDBA},
\ref{f:THUD}, and \ref{f:OKNDA}.

For a specified universal IMF, the
key parameter appears to be the
probability of a region being
active, $\chi$.   The chemical
evolution of the SN thick, thin,
and thick + thin disk,
appear to be characterized by
low values, $\chi=0.32$, 0.13, 0.07,
respectively, while a high value,
$\chi=0.89$, is related
to the chemical evolution of the
bulge, as shown in Table
\ref{t:IMFU}.   In addition, the
chemical evolution of the
SN inner halo
is characterized by
$\chi=0.013$, a value close to
one tenth its
SN thin disk counterpart.   The
corresponding true oxygen yield,
$\hat{p}=0.0058$, is consistent
with $\hat{p}=0.0085\mp0.0035$
deduced from observations (e.g.,
\cite{pra94}).

In conclusion, a single inhomogeneous
model of chemical evolution with
universal IMF and gas neither inhibited
from, nor enhanced in, star formation,
provides acceptable fits to the EDOD
related to both the SN inner halo,
the SN thick disk, the SN thin disk,
the SN thick + thin disk, and the bulge.
In particular, the main part of objects
of uncertain population could belong to
the thin disk, as shown in Fig.\,\ref{f:THUD}.
Conversely, an inhomogeneous model with
considerable amount of gas inhibited from
star formation (H1), used in an earlier
attempt (C07), provides an acceptable fit
to the outer halo.

With regard to the SN thick disk,
the SN thin disk, and the bulge,
both poor and rich samples are
fitted by the same models.   With
regard to the SN halo, the existence
of two different classes of objects
in the (incomplete) related sample,
has to be advocated: a low-metallicity
tail belonging to the outer halo, and
a higher-metallicity body belonging
to the inner halo, each fitted by a
different model, as shown in 
Fig.\,\ref{f:THUD}.  A possible
explanation shall be given below.

\subsection{A physical interpretation
of the model} \label{ss:pim}

The EDODs related to MaK08, MaN08,
MaK08 + MaN08, MaB08, RaK07, RaN07,
RaK07 + RaN07 subsamples, are fitted
to an acceptable extent by models
DB105, DN81, KN6, B,
listed in Table \ref{t:IMFU}.   In
addition, the EDOD related to the
RaH07 subsample is fitted to an
acceptable extent by model H listed
in Table \ref{t:IMFU}, provided the
low-metallicity tail is not considered.
Present-day gas mass fraction and
oxygen abundance related to the SN
disk may be expressed as weighted
means:
\begin{lefteqnarray}
\label{eq:muf}
&& \mu_f=\frac{M_{\rm g}}M=\frac{M_{\rm Kg}+M_{\rm Ng}}{M_{\rm K}+M_{\rm N}}
\nonumber \\
&& \phantom{\mu_f}
=\frac{M_{\rm K}\mu_{\rm Kg}}{M_{\rm K}+M_{\rm N}}+
\frac{M_{\rm N}\mu_{\rm Ng}}{M_{\rm K}+M_{\rm N}}~~; \\
\label{eq:ZOf}
&& (Z_{\rm O})_f=\frac{(M_{\rm O})_{\rm g}}{M_{\rm g}}=
\frac{(M_{\rm O})_{\rm Kg}+(M_{\rm O})_{\rm Ng}}{M_{\rm Kg}+M_{\rm Ng}}
\nonumber \\
&& \phantom{(Z_{\rm O})_f}
=\frac{M_{\rm Kg}(Z_{\rm O})_{\rm Kg}}{M_{\rm Kg}+M_{\rm Ng}}+
\frac{M_{\rm Ng}(Z_{\rm O})_{\rm Ng}}{M_{\rm Kg}+M_{\rm Ng}}~~;\quad
\end{lefteqnarray}
where $M_{\rm g}$ and $M$ are the present gas
and total mass, and the indices, K and N,
denote the thick and the thin disk,
respectively.   For models DB105 and
DN81, the reference values, $M_{\rm K}/
M_{\rm N}=0.1,~1,$ yield $\mu_f=0.13746,$
0.19998, respectively, to be compared to
$\mu_f=0.21244$ related to model KN6,
where the thick and thin disk evolve
as a single system.

Models listed in Table \ref{t:IMFU} are
related to a strictly universal IMF,
in the sense that both the power-law
exponent and the upper and lower stellar
mass limit coincide in all cases and, in
addition, gas is neither inhibited
from, nor enhanced in, star formation,
concerning one model with respect to
the other.   This is a stronger
condition in comparison with earlier
results (C01, C07, C08) based on SN
disk samples biased towards
high-metallicity stars.

For models characterized by a strictly
universal IMF, the probability of a region
being active, $\chi$, is highly relevant.
An inspection of Table \ref{t:IMFU} shows
that $\chi=0.01268,$ 0.07405, for the SN
inner halo and the SN thick + thin disk,
respectively, increasing up to $\chi=
0.12549,$ 0.32460, 0.88717, for the SN
thin disk, the SN thick disk, and the bulge,
respectively.   A physical interpretation
of the above results, in the light of the
model, is that the probability of a region
being active mainly depends on the
environment.

More specifically,
the star formation rate within an
active region is independent of
the environment in the case under
discussion.   On the contrary, a
different global efficiency (which
is proportional to the probability
of a region being active) takes
place in different environments.
Using the values of $\chi$ listed
in Table \ref{t:IMFU}, the following
global efficiency ratios are calculated:
$\chi_{\rm B}/\chi_{\rm H}\approx70;$
$\chi_{\rm B}/\chi_{\rm KN}\approx12;$
$\chi_{\rm B}/\chi_{\rm N}\approx7;$
$\chi_{\rm B}/\chi_{\rm K}\approx3;$
where B, H, K, N, KN denote the bulge,
the SN halo, the SN thick disk, the
SN thin disk, the SN thick + thin
disk, respectively.

Star formation is triggered by tidal
interaction (including merging, which
can be conceived as the strongest
tidal interaction).   Accordingly,
lower $\chi$ values are expected in
environments where tidal
effects are poorly experienced, such
as the inner halo, the thick + thin
disk, and the thin disk.   On the
other hand, larger $\chi$ values are
expected in environments where
considerable tidal effects take
place, such as the thick disk and
the bulge.   According to recent
findings (e.g., \cite{mea08})
the thick disk and the bulge
formation timescales were similar,
of the order of several tenths of
Gyr.   The density is higher at the
end of contraction, which could
maximize tidal effects between
different regions, and then produce
a larger number of active regions
with respect to
environments where the density
is considerably lower (inner halo)
or the chemical evolution timescale
considerably longer (thin disk).

\subsection{Universal IMF and
G-dwarf problem} \label{ss:IMFGD}

A different IMF in bulge and disk
components has been advocated for
explaining the metallicity distribution
in both the Milky Way and Andromeda
bulge (\cite{bma07}), contrary
to what has been deduced from recent
observations performed on SN thick disk,
SN thin disk, and Milky Way bulge stars
(\cite{mea08}).   As pointed
out in an earlier attempt (C08), a
large fraction of Fe production
depends on type Ia supernovae which,
in turn, are thought to be born from
an accreting white dwarf progenitor
in a binary system,
including a red giant outside its Roche lobe.
Accordingly, different IMFs in different
environments could be simulated by
different amounts of binary star
fractions (in particular, SnIa progenitors).
In this view, denser
environments where objects with opposite
velocity vectors are close one to the
other (e.g., the assembling bulge, the
assembling thick disk) should be expected
to exhibit a smaller binary star fraction
(in particular, SnIa progenitors), with
respect to less dense environments where
objects with parallel velocity vectors
are close one to the other (e.g., the
assembled thin disk).

Regions undergoing an early infall phase
(of primordial composition), with the
smaller regions having a more important
infall, have been advocated for explaining
the field star metallicity distribution
in both the halo and dwarf spheroidal
satellites of the Milky Way, where a
G-dwarf problem seems to occur (\cite{pra94}).
Though a G-dwarf problem (i.e.
overabundance of low-metallicity, long-lived
stars predicted by the Simple model with
respect to observations) has been detected
both in bulge-dominated and disk-dominated
galaxies (\cite{hew99}) and is
probably universal (\cite{woa96}),
still it may be explained by a large number
of different ways (e.g., \cite{pap75})
including, among others, unprocessed
gas inflow.   On the other hand, a prompt
initial enrichment from pop.\,III stars
cannot be overlooked and, in particular,
the idea of a coeval formation of pop.\,III
and pop.\,II stars cannot be disregarded
(e.g., \cite{sma09}).   In this
view, the G-dwarf problem in old stellar
systems (e.g., the halo) could find a
natural explanation, if still occurring
within the framework of inhomogeneous
models of chemical evolution.

\subsection{Stellar migration across
the disk} \label{ss:ormi}

Though sample objects considered in the
current paper, with the exception of
bulge stars, are made of SN stars,
still related informations provide
valuable clues for understanding the
formation process of the Galaxy.
The assumption that the EDOD of the
local disk is representative of the
global disk, 
even if in contrast with an inside-out
disk formation, can be considered as a
useful zero-th order approximation.   On
the other hand, nearby stars older than
about 0.2 Gyr come from birth sites which 
span a large range in Galactocentric
distances (e.g., \cite{roa00}).
The orbital diffusion coefficient deduced
from the observed increase of velocity
dispersion with age implies that presently
local stars have suffered a rms azimuthal
drift from about 2 kpc (for an age of 0.2
Gyr) to many Galactic orbits (for an age
of 10 Gyr); for further details refer to
earlier work 
(\cite{wie77}).   Considerable, but smaller,
drift should occur also on the radial
direction (\cite{wia96}; \cite{hay08}).

In this sense, the star formation
rate inferred for nearby stars is a measure
of the global Milky Way star formation rate,
at least at the sun Galactocentric radius
(\cite{roa00}), according to
the estimates of the diffusion coefficient
(e.g., \cite{mea91}).   Star
migration along the equatorial plane is
mainly due to churning and blurring via
interactions between stars and spiral
arms (Sellwood and Binney, 2002; \cite{scb09}).
Diffusion calculations
based on the increase of local velocity
dispersion with age, could be inaccurate
if significant migration occurs, as due to
resonant scattering with transient spiral
arms, which makes local samples considerably
affected by radial migration, according to
$N$-body + smooth particle hydrodynamics
simulations (\cite{roa08}).

As outlined in an earlier attempt (\cite{was93}),
radial flows of the gas
(possibly due to the observed spiral density
waves) or different disk star formation
histories (with time scales comparable to
that of chemical evolution) between the
inner and the outer parts of the Galaxy,
may change the local abundances, but the
overall abundance in the disk should not
be affected as long as there is no gain or
loss of material in the disk.   It can also
be noticed (\cite{was93}) that the
average oxygen abundance in the disk is
roughly the solar value, and can be plausibly
explained by the standard Scalo IMF with
lower and upper star mass limit, $\widetilde
{m}_{\rm mf}=0.1$ and $\widetilde{m}_{\rm Mf}=60$,
respectively, or by a power-law IMF with
exponent, $p$, within the range, $2.35\le
p\le2.9$, and the same $\widetilde{m}_{\rm mf}$ 
and $\widetilde{m}_{\rm Mf}$.   For further
details refer to the parent paper (\cite{was93}).

Inhomogeneous models can describe to some
extent stellar migration across the disk,
provided inflow
(i.e. enhanchement of star formation) is
considered.   As outlined in an earlier
attempt (C07), inhomogeneous models where
star formation is enhanced imply star inflow
with same oxygen abundance as in the existing
gas, which mimics (a special case of) stellar
migration.   Mass conservation no longer
holds, and the effective gas mass fraction
(normalized to the final mass), $\mu_{\rm eff}$,
is expressed in terms of the actual gas mass
fraction (normalized to the initial mass), $\mu$, as:
\begin{equation}
\label{eq:muef}
\mu_{\rm eff}=\frac\mu{1+s_{\rm mig}}~~;
\end{equation}
where $s_{\rm mig}$ is the inflowed star mass
fraction (normalized to the initial mass),
to be interpreted as the net stellar migration
within the volume under consideration, i.e.
the difference between the fractional mass
in stars which have been born elsewhere and
migrated inside, and born inside and migrated
elsewhere, respectively.

In the framework of inhomogeneous models with
enhanced star formation, the fractional mass
in migrated stars is proportional to the
fractional mass in stars generated in situ,
$s_{\rm mig}=\zeta s=\zeta(1-\mu)$, and
Eq.\,(\ref{eq:muef}) may be rewritten as:
\begin{equation}
\label{eq:muez}
\mu=\frac{(1+\zeta)\mu_{\rm eff}}{1+\zeta\mu_{\rm eff}}~~;
\end{equation}
where the effective gas mass fraction, $\mu_{\rm eff}$,
can be deduced from the observations.   For the
solar neighbourhood,
$0.1\appleq\mu_{\rm eff}\appleq0.3$ (\cite{pra95}),
$\zeta\approx1$ from high-resolution $N$-body +
smooth particle hydrodynamics simulations of disk
formation (\cite{roa08}), and
Eq.\,(\ref{eq:muez}) yields $0.18\appleq\mu\appleq
0.46$.

On the other hand, model DN4 listed in
Table \ref{t:KN} produces $\mu\approx0.37$,
$s=1-\mu\approx0.63$ and
and $-D_f=s_{\rm mig}\approx0.47$, which implies
about three quarters of stars were born in situ,
and the remaining quarter is the net effect of stellar
migration.    Accordingly, inhomogeneous models
with enhanced star formation provide an acceptable
description of stellar migration even if, in the
case considered, model DN4 fails in fitting the
SN thin disk EDOD, as shown in Fig.\,\ref{f:OKNDA}
(top right panel).


\subsection{Implications for the formation
of the Galaxy} \label{ss:foga}

According to a number of recent investigations,
the formation of the Galaxy appears to be
characterized by the following main features.
\begin{description}
\item[(i)]
Thick disk stars had a chemical enrichment
similar and coeval to metal-rich halo stars,
[Fe/H]$\approx-1.3$ (\cite{pra00}),
$-1.5<$[Fe/H]$<-0.1$ (\cite{raa07}),
$-1.3<$[Fe/H]$<-0.5$ (\cite{mea08}).
\item[(ii)]
Thick disk stars had a chemical enrichment
similar and coeval to metal-poor bulge stars,
[Fe/H]$\appleq-0.4$ (\cite{pra00}),
$-1.3<$[Fe/H]$\le-0.5$ (\cite{mea08}).
\item[(iii)]
The thick disk low-metallicity tail ends near
[Fe/H]$=-2.2$ (\cite{chb00}).
\item[(iv)]
The bulge formation timescale is about 1 Gyr
at most (\cite{zoa08}).
\item[(v)]
The halo is made of two broadly overlapping
structural components: an inner halo with
axis ratio close to 0.6, modest net prograde
rotation, and peak metallicity at [Fe/H]$=-
1.6$; and an outer halo with nearly spherical
shape, net retrograde rotation, and peak
metallicity at [Fe/H]$=-2.2$ (\cite{caa07}).
\item[(vi)]
The mean rotational velocities are close to
zero or negative for the outer halo $(R\appgeq12$\,kpc),
which may correspond to the radial limit
of the rapidly rotating thick disk component
(\cite{chb00}; \cite{caa07}).
\item[(vii)]
A linear dependence of mean rotational
velocities on the metallicity for [Fe/H]$
\appgeq-1.6$, appears to be consistent
with the contraction
of the inner halo (\cite{chb00})
after an earlier assembling phase.
Interestingly, the inner halo metallicity peaks
around [Fe/H]$\approx-1.6$ (e.g., \cite{ryn91};
\cite{caa07}; \cite{pra08}).
\item[(viii)]
The distribution of specific angular
momentum appears to be consistent with
both a halo-bulge and a thick disk-thin
disk collapse, but not with a spheroid-disk
collapse (\cite{wyg92}; \cite{ibg95}).
\end{description}

At this point, the question arises if the
above mentioned results are a natural
consequence of the currently favoured
theory of galaxy formation based on
hierarchical assembly of cold dark
matter haloes in presence of both
baryonic matter and quintessence
(e.g., \cite{mov04};
\cite{hob05}).   In the
framework of QCDM models, the initial
density perturbations in the early
universe have larger amplitudes on
smaller scales, which makes low-mass
density perturbations turn around and
recollapse first.   Accordingly, star formation
is expected to start therein.   For
further details refer to earlier
attempts related to CDM scenarios
(e.g., \cite{cai97}; \cite{chb00}).

Low-mass density perturbations
lying within a large-mass density
perturbation, may be considered as
secondary peaks bound to a primary
peak (e.g., \cite{ryd88}).   In
particular, objects existing at
present with Galactic mass are
likely built up by the merging
and accretion of smaller, less
massive progenitors.   It may
safely be thought that these
progenitors - at least in the
denser regions of the universe -
allow baryons to concentrate in
their cores.   This process
takes place as an inevitable
consequence of energy dissipation
via radiative cooling and
anelastic collisions, which
does not carry significant
amount of angular momentum.

Stellar systems (and sufficiently
clumpy gas clouds) moving along
highly eccentric and/or highly
inclinated orbits, make angular
momentum transport possible via
quadrupole interaction.   More
specifically, angular momentum
is transferred from the orbital
motion of the tighly bound cores
to the loosely bound outer halo
particles as the cores sink into
the centre of the forming halo.
This loss of angular momentum by
the cores, due to dynamical
friction, will lead to a low
specific angular momentum in
the assembled galaxy, as seen
in ellipticals.   For further
details refer to earlier attempts
(\cite{zua88}; \cite{bec00}).
On the other
hand, stellar systems formed
from the gas heated in the course
of merging and accretion, when a
substantial fraction of peculiar
energy has been dissipated, would
maintain their specific angular
momentum, as seen in disk galaxies.

In this view, the inner
halo should have a flattened density
distribution with a finite prograde
rotation, conformly to observations
(\cite{chb00}; \cite{caa07}), and the
angular momentum distribution may
be similar to its bulge counterpart
(\cite{wyg92}).   The thick
disk could have been accreted from the
external part of the proto-Galaxy (e.g.,
\cite{sla97}), and its
angular momentum distribution may be
similar to its thin disk counterpart
(\cite{ibg95}).   In other
words, the inner halo and the bulge
could take origin from the denser
internal region of the initial density
perturbation, where smaller spin growth
via tidal interactions with neighbouring
objects occurs, while the less dense
external region, characterized by a
larger spin growth, could be the
progenitor of the thick and thin
disk.   The outer halo could be built
up via accretion of smaller fragments
or satellites even if, initially, bound
to different peaks.

The building blocks
of each subsystem may be conceived in a
twofold manner, as lower mass overdensities
which first decouple from the universe
(progenitors of compact dwarf galaxies)
from the cosmological side, and upper
mass gas media which form stars (molecular
clouds or proto-globular clusters) from
the astrophysical side.

In the framework of inhomogeneous
models of chemical evolution considered
in the current paper, star formation
takes place within a fixed fraction of
identical regions.   At the end of any
step, the enriched gas of active regions
is homogeneously mixed with the unprocessed
gas of quiescent regions, and a new set of
identical regions is produced, which makes
the beginning of the subsequent step.
Accordingly, the chemical enrichment
history in different environments, is
expected to be similar and even coeval for
a strictly universal IMF.   In fact, the
total number of steps is an input parameter,
but the duration of a step can freely be
assigned (C00, C01, C07).
For fiducial values of mass and radius
related to the proto-bulge $(M={\rm M_{10}},
\,R=6\,{\rm kpc})$ and the proto-inner halo
+ disk
$(M=6\,{\rm M_{10}},\,R=15\,{\rm kpc})$, the
free-fall time is 0.08-0.12 Gyr, a few
times lower than the bulge formation
timescale inferred from observations
(\cite{zoa08}).

Different environments could imply a
different probability of a region
being active i.e. a different ratio
of active to active + quiescent regions
at any step and, in turn, different
global efficiencies of the star formation
rate.   In this sense, chemical
evolution could be influenced by dynamical
evolution: for instance, a higher number
of active regions corresponds to environments
where tidal effects between neighbouring
regions are strong (e.g., the bulge) and
vice versa (e.g., the halo).

Models implying large amount of gas inhibited
from star formation (H1 in the current paper)
succeed in reproducing (incomplete) data related
to the outer halo population.   To this respect,
some considerations seem useful provided the
fit shall be confirmed using richer samples.
Even if mass conservation formally holds in
models under discussion, the chemical evolution
acts as gas inhibited from star formation and
a fraction of gas synthesized in short-lived
stars and returned to the interstellar medium,
were lost from the outer halo and presumably
accreted by the proto-Galaxy.

Using an outer halo star mass fraction, $s_f=
0.13$, inhibited (i.e. lost) gas mass
fraction, $D_f=0.87$, as listed in Table
\ref{t:DHB} for model H1, and assuming a 
stellar outer halo upper mass limit equal to $10^9
{\rm m}_\odot$, the total mass lost from
the outer halo cannot exceed $6.7\,10^9
{\rm m}_\odot$.   The total mass of the
proto-Galaxy is about ten times larger,
which makes little influence on the
related chemical evolution.   In
conclusion, mass loss from active regions
makes a different oxygen enrichment with
respect to environments (inner halo, thick
disk, thin disk, bulge) where mass conservation
holds to a better extent.

The observed abundance distributions
used in the current paper provide
additional support to points (i),
(ii), (iii), (v), listed above,
where points (vi) and (vii) may be
considered as complementary to (v).
The comparison with model predictions,
related to oxygen, is also consistent
with the above mentioned points: the
SN thick disk, the SN inner halo, and
the bulge had similar and coeval
oxygen enrichment within the common
abundance range.   On the other hand,
a different model with gas inhibited
from star formation has to be used
for the outer halo, which is consistent
with points (v), (vi), and (vii).

Keeping in mind point (iv), the above
results imply a short formation timescale
for both the (metal-rich) inner halo,
the thick disk, the thin disk, and the
bulge.   Oxygen enrichment within a
common range in different subsystems
seems to be slightly dependent on
dynamical parameters, according to
point (viii).   If any two subsystems
formed on comparable timescales, as
well as had a strictly universal IMF
and star formation histories, the
abundance pattern would be similar
even if the two populations lack a
physical connection.

An idea of how galaxies did form is
offered by celestial bodies which are
presently assembling i.e. cluster of
galaxies.   Similarly, individual
proto-galaxies may be conceived as
made of subunits (Still visible dwarf
spheroidals?  Subsequently dead dwarf
spheroidals with surviving nuclei in
form of globular clusters?   Globular
clusters?) which are virializing in
the inner volume but are still falling
in the outer volume, while a central
structure (the proto-bulge) is growing
up.

As outlined in an earlier attempt
(\cite{chb00}), more elaborate
numerical modelling of the formation
of large spiral galaxies, such as the
Milky Way, is also needed in order to
clarify the physical processes that
lead to the currently observed dynamics
and structure of the spheroid and disk
components.   It is of particular
importance to model and understand the
chemo-dynamical evolution of the system
of sub-galactic fragments in the course
of the Galaxy assembling (e.g., \cite{pra94}).
A fundamental
understanding of the formation and the
evolution of the Milky Way, is expected
to provide additional insights on the
formation and the evolution of disk-type
galaxies.

\section{Conclusion} \label{s:conc}

The main results of the present attempt
may be summarized as follows.
\begin{description}
\item[(1)]
The empirical differential oxygen abundance
distribution (EDOD) is determined
from recent data involving different subsamples
related to the solar neighbourhood (SN) halo,
thick disk, thin disk, and to the bulge (\cite{raa07});
\cite{mea08}).
\item[(2)]
The EDOD related to different subsamples
(including less recent ones)
and/or different populations, show qualitative
agreement above a threshold in iron abundance,
[Fe/H]$>-$1, where the data in richer
subsamples are statistically significant.
The sole exception concerns halo stars,
possibly due to subsample incompleteness
(\cite{raa07}) and/or different halo
populations (\cite{caa07}).
\item[(3)]
The SN thick + thin disk EDOD is
deduced from its thick disk and thin disk
counterparts, for different choices of
the thick disk to thin disk mass ratio
within the range, 0.1-0.9.   The related
effect is comparable to
the error calculated for the EDOD.
\item[(4)]
Inhomogeneous models with strictly
universal initial mass function (IMF),
implying gas is neither inhibited from,
nor enhanced in, forming stars at
different rates with respect to a selected
reference case, provide
an acceptable fit to the EDMD related to
different environments.   To this aim, a
SN thin disk low-metallicity tail (down
to $\phi\approx0.2$) has to be assumed,
and a single oxygen overabundant sample
star must be
disregarded as a peculiar object and/or
affected by undetected systematic error.
The existence of a strictly universal
IMF implies a similar chemical enrichment
within active regions placed in different
environments which, on the other hand,
exhibit increasing probability of a region
being active, and then an increasing
global efficiency of the star
formation rate, passing from the SN halo to
the SN thin disk, the SN thick disk, and
the bulge.  The evolution of the SN thick
+ thin disk as a whole (\cite{hay08}),
computed using inhomogeneous models, is
also consistent with the related EDMD.
\item[(5)]
A special case of stellar migration
across the disk (inflowing stars with equal
oxygen abundance with respect to the existing
gas) can be described by inhomogeneous
models with enhanced star formation.
In particular, values within the observed
range in SN
gas mass fraction may allow a fraction
of present SN stars were born in situ, and a
comparable fraction is due to the net effect
of stellar
migration, according to recent results
based on high-resolution $N$-body + smooth
particle hydrodynamics simulations (\cite{roa08}).
\end{description}

\acknowledgments
The authors are grateful to an anonymous referee
for critical comments, which made substantial
improvement of an earlier version of the paper.
Thanks are due to R. Ro\v{s}kar for pointing
our attention to a recent paper (\cite{roa08}),
quoted in the text.

\appendix
\section*{Appendix}

\section{Caption of symbols}\label{a:simbo}

To help the reader, the meaning of the parameters
listed in the Tables throughout the paper, is
explained below.  The end of evolution is
denoted by the index, $f$.   For further details
refer to parent papers (C00, C01, C07, C08).

\begin{description}
\item[$Z_{\rm O}$]
Mass oxygen abundance.
\item[$\phi$]
Mass oxygen abundance normalized to the solar value.
\item[$\Delta\phi$]
Mass oxygen abundance (normalized to the solar value)
bin, related to a selected [O/H] bin.
\item[$\phi^\mp$]
Upper (plus) and lower (minus) normalized oxygen
abundance limit, related to an assigned normalized
oxygen abundance bin, $\Delta\phi$.
\item[$\Delta^\mp\phi$]
Right (plus) and left (minus) normalized oxygen
abundance interval semiamplitude related
to an assigned normalized oxygen abundance bin,
$\Delta\phi$.
\item[$\Delta^\ast\phi_{\rm R}^\prime$]
Newly synthesised oxygen gas mass fraction (normalized
to the solar value) within an active region, in
inhomogeneous models of chemical evolution, at the
end of a step.
\item[$\Delta^\ast\phi$]
Newly synthesised oxygen gas mass fraction (normalized
to the solar value) within the whole system, in
inhomogeneous models of chemical evolution, at the
end of a step during the expected evolution.
\item[$\psi$\phantom{en}]
Empirical differential oxygen abundance distribution
(EDOD). \\
\phantom{$\psi$i}
Theoretical differential oxygen abundance distribution
(TDOD).
\item[$\psi^\mp$]
Upper (plus) and lower (minus) EDOD limit related
to an assigned normalized oxygen abundance bin,
$\Delta\phi$.
\item[$\Delta^\mp\psi$]
Top (plus) and bottom (minus) EDOD interval
semiamplitude related
to an assigned normalized oxygen abundance bin,
$\Delta\phi$.
\item[$\psi_0$\phantom{en}]
Theoretical differential oxygen abundance distribution
(exact value) at the starting point of the expected
evolution (C01, C07).
\item[$\psi_1(40)$\phantom{en}]
Theoretical differential oxygen abundance distribution
(exact value) at the ending point of the first step of
the expected evolution [C01, C07, Eq.\,(40)].
\item[$\psi_1(41)$\phantom{en}]
Theoretical differential oxygen abundance distribution
(second-order approximation) at the ending point of the
first step of the evolution in the general case [C01,
C07, Eq.\,(41)].
\item[$\mu_{\rm R}^\prime$]
Gas mass fraction which allows star formation within
an active region at the end of a step.
\item[$D_{\rm R}^\prime$]
Gas mass fraction which inhibits star formation
within an active region at the end of a step.
Negative values denote enhancement of star
formation due to gas inflow.
\item[$s_{\rm R}^\prime$]
Long-lived star mass fraction within an active
region at the end of a step.
\item[$q$]
Effective gas mass fraction within a
region at the end of a step.
\item[$\mu$]
Gas mass fraction which allows star formation.
\item[$D$]
Gas mass fraction which inhibits star formation.
Negative values denote enhancement of star
formation due to gas inflow.
\item[$s$]
Long-lived star mass fraction.
\item[$\alpha$]
Mass fraction of a star generation which remains
locked up in long-lived stars and stellar remnants.
\item[$\kappa$]
Ratio of gas mass fraction which inhibits star
formation to long-lived star and stellar remnant mass fraction.
Negative values denote enhancement of star
formation due to gas inflow.
\item[$m_{\rm mf}$]
Lower mass limit of long-lived stars.
\item[$\hat{p}$]
True oxygen yield.
\item[$\hat{p}^\prime$]
Effective oxygen yield related to both inhomogeneous
star formation and inhibited or enhanced star formation.
\item[$\hat{p}^{\prime\prime}$]
Effective oxygen yield related to inhibited or
enhanced star formation.
\item[$\chi$]
Probability of a region being active.
\end{description}

\section{Table \ref{t:KN}}\label{a:tab}

Table \ref{t:KN} is exceedingly large in the text, and for this
reason it is broken to be completely accessible.

%
\begin{table*}
\caption[par]{Table \ref{t:KN}, left.}
\begin{center}
\begin{tabular}{lllll}
\multicolumn{1}{c}{} &
\multicolumn{1}{c}{DB12} & \multicolumn{1}{c}{DB13} &
\multicolumn{1}{c}{DB14} & \multicolumn{1}{c}{DB15} \\
\noalign{\smallskip}
\hline\noalign{\smallskip}
$\mu_{\rm R}^\prime$               & 1.3243~E$-$1 & 1.2295~E$-$1 & 1.1258~E$-$1 & 1.0304~E$-$1 \\
$q$                                & 6.1771~E$-$1 & 5.5533~E$-$1 & 4.9941~E$-$1 & 4.4924~E$-$1 \\
$\chi$                             & 4.4156~E$-$1 & 5.0700~E$-$1 & 5.6410~E$-$1 & 6.1403~E$-$1 \\
$\hat{p}/(Z_{\rm O})_\odot$        & 9.0478~E$-$1 & 8.3518~E$-$1 & 7.7553~E$-$1 & 7.2382~E$-$1 \\
$\hat{p}^\prime/(Z_{\rm O})_\odot$ & 3.6191~E$-$1 & 3.3407~E$-$1 & 3.1021~E$-$1 & 2.8953~E$-$1 \\
$\mu_f$                            & 1.4559~E$-$1 & 9.5108~E$-$2 & 6.2204~E$-$2 & 4.0731~E$-$2 \\
$\alpha_{2.9}$                     & 6.9505~E$-$1 & 7.1175~E$-$1 & 7.2671~E$-$1 & 7.4020~E$-$1 \\
$\alpha_{2.35}$                    & 8.6951~E$-$1 & 8.7833~E$-$1 & 8.8803~E$-$1 & 8.9281~E$-$1 \\
$(\widetilde{m}_{\rm mf})_{2.9}$       & 4.0227~E$-$1 & 3.7813~E$-$1 & 3.5660~E$-$1 & 3.3728~E$-$1 \\
$(\widetilde{m}_{\rm mf})_{2.35}$      & 1.1305~E$-$2 & 9.3463~E$-$3 & 7.8194~E$-$3 & 6.6108~E$-$3 \\
$\Delta^\ast\phi$                  & 1.7435~E$-$1 & 1.9650~E$-$1 & 2.1539~E$-$1 & 2.3168~E$-$1 \\
$\Delta^\ast\phi_{\rm R}^\prime$   & 1.8170~E$-$0 & 1.7505~E$-$0 & 1.6938~E$-$0 & 1.6450~E$-$0 \\
$-\psi_o$                          & 2.4322~E$-$1 & 1.7337~E$-$1 & 1.1035~E$-$1 & 5.3384~E$-$2 \\
$-\psi_1(40)$                      & 2.8439~E$-$1 & 2.2346~E$-$1 & 1.6926~E$-$1 & 1.2104~E$-$1 \\
$-\psi_1(41)$                      & 2.8506~E$-$1 & 2.2446~E$-$1 & 1.7066~E$-$1 & 1.2289~E$-$1 \\
                                   &              &              &              &              \\
$\kappa$                           & 1.4286~E$-$1 & 2.3810~E$-$1 & 3.3333~E$-$1 & 4.2857~E$-$1 \\
$s_{\rm R}^\prime$                 & 7.5755~E$-$1 & 7.0839~E$-$1 & 6.6557~E$-$1 & 6.2787~E$-$1 \\
$D_{\rm R}^\prime$                 & 1.0822~E$-$1 & 1.6866~E$-$1 & 2.2186~E$-$1 & 2.6909~E$-$1 \\
$\overline{\phi}$                  & 7.8307~E$-$1 & 7.4978~E$-$1 & 7.2065~E$-$1 & 6.9485~E$-$1 \\
$s_f$                              & 7.4761~E$-$1 & 7.3087~E$-$1 & 7.0335~E$-$1 & 6.7149~E$-$1 \\
$D_f$                              & 1.0680~E$-$1 & 1.7402~E$-$1 & 2.3445~E$-$1 & 2.8778~E$-$1 \\
\noalign{\smallskip}
\hline
\end{tabular}
\end{center}
\end{table*}
\begin{table*}
\caption[par]{Table \ref{t:KN}, right.}
\begin{center}
\begin{tabular}{llrrrl}
\multicolumn{1}{c}{} &
\multicolumn{1}{c}{DN}  &
\multicolumn{1}{c}{DN4} &
\multicolumn{1}{c}{DN5} &
\multicolumn{1}{c}{DN6} & \multicolumn{1}{c}{DN7}  \\
\noalign{\smallskip}
\hline\noalign{\smallskip}
$\mu_{\rm R}^\prime$               & 1.6663~E$-$1 &    3.7477~E$-$1 &    3.6662~E$-$1 &     3.5864~E$-$1 & 3.5083~E$-$1 \\
$q$                                & 8.0550~E$-$1 &    9.6146~E$-$1 &    9.3741~E$-$1 &     9.1396~E$-$1 & 8.9109~E$-$1 \\
$\chi$                             & 2.3339~E$-$1 &    6.1635~E$-$2 &    9.8821~E$-$2 &     1.3416~E$-$1 & 1.6776~E$-$1 \\
$\hat{p}/(Z_{\rm O})_\odot$        & 7.2382~E$-$1 &    1.8096~E$-$0 &    1.4476~E$-$0 &     1.2064~E$-$0 & 1.0340~E$-$0 \\
$\hat{p}^\prime/(Z_{\rm O})_\odot$ & 2.8953~E$-$1 &    1.0857~E$-$0 &    8.6859~E$-$1 &     7.2382~E$-$1 & 6.2042~E$-$1 \\
$\mu_f$                            & 4.4842~E$-$3 &    3.7439~E$-$1 &    1.9872~E$-$1 &     1.0548~E$-$1 & 5.5988~E$-$2 \\
$\alpha_{2.9}$                     & 7.4020~E$-$1 &    5.3263~E$-$1 &    5.8755~E$-$1 &     6.3092~E$-$1 & 6.6604~E$-$1 \\
$\alpha_{2.35}$                    & 8.9281~E$-$1 &    7.6914~E$-$1 &    8.0638~E$-$1 &     8.3327~E$-$1 & 8.5360~E$-$1 \\
$(\widetilde{m}_{\rm mf})_{2.9}$       & 3.3728~E$-$1 &    6.4228~E$-$1 &    5.6022~E$-$1 &     4.9602~E$-$1 & 4.4449~E$-$1 \\
$(\widetilde{m}_{\rm mf})_{2.35}$      & 6.6108~E$-$3 &    5.1835~E$-$2 &    3.2618~E$-$2 &     2.1898~E$-$2 & 1.5436~E$-$2 \\
$\Delta^\ast\phi$                  & 6.2622~E$-$2 &    4.2667~E$-$2 &    5.6142~E$-$2 &     6.5123~E$-$2 & 7.1537~E$-$2 \\
$\Delta^\ast\phi_{\rm R}^\prime$   & 1.2971~E$-$0 &    1.7760~E$-$0 &    1.4526~E$-$0 &     1.2370~E$-$0 & 1.0831~E$-$0 \\
$-\psi_o$                          & 4.8961~E$-$2 &    1.2641~E$-$0 &    1.0696~E$-$0 &     9.0546~E$-$1 & 7.6482~E$-$1 \\
$-\psi_1(40)$                      & 5.0826~E$-$1 &    1.2692~E$-$0 &    1.0780~E$-$0 &     9.1713~E$-$1 & 7.7976~E$-$1 \\
$-\psi_1(41)$                      & 5.0839~E$-$1 &    1.2692~E$-$0 &    1.0780~E$-$0 &     9.1718~E$-$1 & 7.7985~E$-$1 \\
                                   &              &                 &                 &                  &              \\
$\kappa$                           & 4.2857~E$-$1 & $-$4.2857~E$-$1 & $-$2.8571~E$-$1 &  $-$1.4286~E$-$1 & 0.0000~E$-$0 \\
$s_{\rm R}^\prime$                 & 5.8336~E$-$1 &    1.0942~E$-$0 &    8.8673~E$-$1 &     7.4825~E$-$1 & 6.4917~E$-$1 \\
$D_{\rm R}^\prime$                 & 2.5001~E$-$1 & $-$4.6892~E$-$1 & $-$2.5335~E$-$1 &  $-$1.0689~E$-$1 & 0.0000~E$-$0 \\
$\overline{\phi}$                  & 8.6448~E$-$1 &    1.1450~E$-$0 &    1.0068~E$-$0 &     9.1462~E$-$1 & 8.4870~E$-$1 \\
$s_f$                              & 6.9686~E$-$1 &    1.0948~E$-$0 &    1.1218~E$-$0 &     1.0436~E$-$0 & 9.4401~E$-$1 \\
$D_f$                              & 2.9865~E$-$1 & $-$4.6921~E$-$1 & $-$3.2051~E$-$1 &  $-$1.4909~E$-$1 & 0.0000~E$-$0 \\
\noalign{\smallskip}
\hline
\end{tabular}
\end{center}
\end{table*}


\begin{thebibliography}{}
%
%
\bibitem[\protect\citeauthoryear{Abia and Rebolo}{1989}]{abr89}  
Abia,~C., Rebolo,~R.: \apj~347, 186 (1989)
\bibitem[\protect\citeauthoryear{Allende-Prieto et al.} {2001}]{apa01}   
Allende-Prieto,~C., Lambert,~D.L.,~Asplund M.: \apjl~556, L63 (2001)
\bibitem[\protect\citeauthoryear{Asplund et al.} {2004}]{asa04}   
Asplund,~M., Grevesse,~N., Sauval,~A.J., et al.: \aap~417,
           751 (2004) 
\bibitem[\protect\citeauthoryear{Ballero et al.} {2007a}]{bka07}  
Ballero,~S.K., Kroupa,~P., Matteucci,~F.: \aap~467, 117 (2007a)
\bibitem[\protect\citeauthoryear{Ballero et al.} {2007b}]{bma07}  
Ballero,~S.K., Matteucci,~F., Origlia,~L., Rich,~R.M.: 
           \aap~467, 123 (2007b)
\bibitem[\protect\citeauthoryear{Barbuy}{1988}]{bar88}  
Barbuy,~B.: \aap~191, 121 (1988) 
\bibitem[\protect\citeauthoryear{Barbuy et al.} {2001}]{baa01}  
Barbuy,~B., Nissen,~P.E.,
Peterson,~R., Spite,~F. (eds.), Proceedings of Oxygen
abundances in old stars and implications for nucleosynthesis and cosmology
(IAU Joint Discussion 8): \nar~45, 509 (2001) 
\bibitem[\protect\citeauthoryear{Bekki and Chiba}{2000}]{bec00}  
Bekki,~K., Chiba,~M.: \apjl~534, L89 (2000) 
\bibitem[\protect\citeauthoryear{Caffau et al.} {2008}]{caa08}  
Caffau,~E., Ludwig,~H.-G., Steffen,~M., et al.: \aap~488,
           1031 (2008) 
\bibitem[\protect\citeauthoryear{Caimmi}{1997}]{cai97}  
Caimmi,~R.: Astron. Nachr. 318, 339 (1997) 
\bibitem[\protect\citeauthoryear{Caimmi}{2000}]{cai00}
Caimmi,~R.: Astron. Nachr. 321, 323 (2000) (C00) 
\bibitem[\protect\citeauthoryear{Caimmi}{2001a}]{cai01a}  
Caimmi,~R.: Astron. Nachr. 322, 65 (2001a) (C00, erratum)
\bibitem[\protect\citeauthoryear{Caimmi}{2001b}]{cai01b}  
Caimmi,~R.: Astron. Nachr. 322, 241 (2001b) (C01)
\bibitem[\protect\citeauthoryear{Caimmi}{2007}]{cai07}  
Caimmi,~R.: \na~12, 289 (2007) (C07)
\bibitem[\protect\citeauthoryear{Caimmi}{2008}]{cai08}  
Caimmi,~R.: \na~13, 314 (2008) (C08)
\bibitem[\protect\citeauthoryear{Carollo et al.}{2007}]{caa07}  
Carollo,~D., Beers,~T.C.Lee,~Y.S., et al.: \nat~318, 1020 (2007) 
\bibitem[\protect\citeauthoryear{Carretta et al.} {2000}]{caa00}  
Carretta,~E., Gratton,~R.G.,  Sneden,~C.: \aap~356, 238 (2000)
\bibitem[\protect\citeauthoryear{Centeno and Socas-Navarro}{2008}]{csn08}  
Centeno,~R., Socas-Navarro,~H.: \apjl~682, L61 (2008)
\bibitem[\protect\citeauthoryear{Chiba and Beers}{2000}]{chb00}  
Chiba,~M., Beers,~T.C.: \aj~119, 2843 (2000) 
\bibitem[\protect\citeauthoryear{Fuhrmann}{2008}]{fuh08}  
Fuhrmann,~K.: \mnras~384, 173 (2008)
\bibitem[\protect\citeauthoryear{Gratton et al.} {2000}]{gra00}  
Gratton,~R.G., Carretta,~E., Matteucci,~F., Sneden,~C.: \aap~358, 671 (2000)
\bibitem[\protect\citeauthoryear{Gray}{2005}]{gra05}  
Gray,~D.F.: The Observation and Analysis of Stellar
           Photospheres, Cambridge University Press, UK (2005)
\bibitem[\protect\citeauthoryear{Haywood}{2001}]{hay01}  
Haywood,~M.: \mnras~325, 1365 (2001) 
\bibitem[\protect\citeauthoryear{Haywood}{2006}]{hay06}  
Haywood,~M.: \mnras~371, 1760 (2006) 
\bibitem[\protect\citeauthoryear{Haywood}{2008}]{hay08}  
Haywood,~M.: \mnras~388, 1175 (2008) 
\bibitem[\protect\citeauthoryear{Henry and Worthey}{1999}]{hew99}  
Henry,~R.B.C., Worthey,~G.: PASP 111, 919 (1999) 
\bibitem[\protect\citeauthoryear{Holmberg et al.}{2007}]{hoa07}  
Holmberg,~J., Nordstr\"om,~B., Andersen,~J.: \aap~475, 519 (2007)
\bibitem[\protect\citeauthoryear{Horellou and Berge}{2005}]{hob05}  
Horellou,~C., Berge,~J.: \mnras~360, 1393 (2005) 
\bibitem[\protect\citeauthoryear{Ibata and Gilmore}{1995}]{ibg95}  
Ibata,~R.A., Gilmore,~G.F.: \mnras~275, 605 (1995) 
\bibitem[\protect\citeauthoryear{Israelian et al.} {2001}]{isa01}  
Israelian,~G., Rebolo,~R., Garcia-Lopez,~R.J., 
Bonifacio,~P., Molaro,~P., Basri,~G., Shchukina,~N.:
\apj~551, 833 (2001)
\bibitem[\protect\citeauthoryear{Ivezic et al.} {2008}]{iva08}  
Ivezic,~Z., Sesar,~B., Juric,~M., et al.: \apj~684, 287 (2008) 
\bibitem[\protect\citeauthoryear{J{\o}rgensen}{2000}]{jor00}  
J{\o}rgensen, B.R., 2000. \aap~363, 947.
\bibitem[\protect\citeauthoryear{Landi et al.} {2007}]{laa07}  
Landi,~E., Feldman,~U., Doschek,~G.A.: \apj~659, 743 ( 2007) 
\bibitem[\protect\citeauthoryear{Malinie et al.} {1993}]{maa93}  
Malinie, G., Hartmann, D.H.,
           Clayton, D.D., Mathews, G.J.: \apj~413, 633 (1993)
\bibitem[\protect\citeauthoryear{Melendez}{2004}]{mel04}  
Melendez,~J.: \apj~615, 1042 (2004) 
\bibitem[\protect\citeauthoryear{Melendez et al.} {2008}]{mea08}  
Melendez,~J., Asplund,~M., Alves-Brito,~A., et al.: \aap~484, L21 (2008) 
\bibitem[\protect\citeauthoryear{Meusinger et al.} {1991}]{mea91}  
Meusinger,~H., Stecklum,~B., Reimann,~H.-G.:
\aap~245, 57 (1991) 
\bibitem[\protect\citeauthoryear{Mota and van de Bruck}{2004}]{mov04}  
Mota,~D.F., van de Bruck,~C.: \aap~421, 71 (2004)
\bibitem[\protect\citeauthoryear{Nordstr\"om et al.}{2004}]{noa04}
Nordstr\"om,~B., Mayor,~M., Andersen,~J., et al.: \aap~418, 989 (2004)
\bibitem[\protect\citeauthoryear{Norris}{1987}]{nor87}  
Norris,~J.E.: \apjl~314, L39 (1987)
\bibitem[\protect\citeauthoryear{Pagel}{1989}]{pag89}  
Pagel,~B.E.J.: The G-dwarf Problem and Radio-active
           Cosmochronology. In: Beckman J.E., Pagel B.E.J. (eds.)
           Evolutionary Phenomena in Galaxies, Cambridge Univ. Press, 
           p.\,201 (1989) 
\bibitem[\protect\citeauthoryear{Pagel and Patchett}{1975}]{pap75}  
Pagel,~B.E.J., Patchett,~B.E.: \mnras~172, 13 (1975)
\bibitem[\protect\citeauthoryear{Prantzos}{1994}]{pra94}  
Prantzos,~N.: \aap~284, 477 (1994)
\bibitem[\protect\citeauthoryear{Prantzos}{2008}]{pra08}  
Prantzos,~N.: EAS Publications Series, vol.\,32, pp.\,311-356 (2008) 
\bibitem[\protect\citeauthoryear{Prantzos and Aubert} {1995}]{pra95}  
Prantzos,~N., Aubert,~O.: \aap~302, 69 (1995) 
\bibitem[\protect\citeauthoryear{Prochaska et al.} {2000}]{pra00}  
Prochaska,~J.X., Naumov,~S.O., Carney,~B.W., 
McWilliam,~A., Wolfe,~A.M.:
\aj~120, 2513 (2000)
\bibitem[\protect\citeauthoryear{Ramirez et al.} {2007}]{raa07}  
Ramirez,~I., Allende Prieto,~C., Lambert,~D.L.: \aap~465,
           271 (2007)
\bibitem[\protect\citeauthoryear{Rocha-Pinto and Maciel}{1996}]{rom96}  
Rocha-Pinto,~H.J., Maciel,~W.J.: \mnras~279, 447 (1996) 
\bibitem[\protect\citeauthoryear{Rocha-Pinto et al.} {2000}]{roa00}  
Rocha-Pinto,~H.J., Maciel,~W.J., Scalo,~J., Flynn,~C.: \aap~358, 850 (2000) 
\bibitem[\protect\citeauthoryear{Ro\v{s}kar et al.} {2008}]{roa08}  
Ro\v{s}kar,~R., Debattista,~V.P., Quinn,~T.R., et al.: 
           \apjl~684, L79 (2008)
\bibitem[\protect\citeauthoryear{Ryan and Norris}{1991}]{ryn91}  
Ryan,~S.G., Norris,~J.E.: \aj~101, 1865 (1991)
\bibitem[\protect\citeauthoryear{Ryden}{1988}]{ryd88}  
Ryden,~B.S.: \apj~329, 589 (1988)
\bibitem[\protect\citeauthoryear{Sadler et al.} {1996}]{saa96}  
Sadler,~E.M., Rich,~R.M., Terndrup,~D.M.: \aj~112, 171 (1996)
\bibitem[\protect\citeauthoryear{Sch\"orck et al.} {2008}]{sca08}  
Sch\"orck,~T., Christlieb,~N, Cohen,~J.G., et al.: 
           Arxiv:0809.1172v1 (2008)
\bibitem[\protect\citeauthoryear{Sch\"onrich and Binney}{2009}]{scb09}  
Sch\"onrich,~R., Binney,~J.: \mnras~  in press,
(Arxiv:0809.3006v1 [astro-ph], 2008) (2009)
\bibitem[\protect\citeauthoryear{Sellwood and Binney}{2002}]{seb02}  
Sellwood,~J.A., Binney,~J.: \mnras~336, 785 (2002)
\bibitem[\protect\citeauthoryear{Smith et al.} {2008}]{sma09}  
Smith,~B.D., Tuck,~M.J., Sigurdsson,~S., et al.: \apj~691, 441 (2009) 
\bibitem[\protect\citeauthoryear{Socas-Navarro and Norton}{2007}]{snn07}  
Socas-Navarro,~H., Norton,~A.A.: \apjl~660, L153 (2007)
\bibitem[\protect\citeauthoryear{Sofia and Meyer}{2001}]{som01}  
Sofia,~U.J., Meyer,~P.M.: \apjl~554, L221 (2001)
\bibitem[\protect\citeauthoryear{Sommer-Larsen et al.} {1997}]{sla97}  
Sommer-Larsen,~J., Beers,~T.C., Flynn,~C., et al.: \apj~481,
           775 (1997) 
\bibitem[\protect\citeauthoryear{Takada-Hidai et al.} {2001}]{taa01}  
Takada-Hidai,~M., Takeda,~Y., Sato,~S., Sargent,~W.L., Lu,~L., Barlow,~T.,
Jugeku,~J.: \nar~45, 549 (2001)
\bibitem[\protect\citeauthoryear{Tinsley}{1980}]{tin80}  
Tinsley,~B.M.: \fcp~5, 287 (1980)
\bibitem[\protect\citeauthoryear{Wang and Silk}{1993}]{was93}  
Wang,~B., Silk,~J.: \apj~406, 580 (1993)
\bibitem[\protect\citeauthoryear{Wielen}{1977}]{wie77}  
Wielen, R. \aap~60, 263 (1977)
\bibitem[\protect\citeauthoryear{Wielen et al.}{1996}]{wia96}  
Wielen,~R., Fuchs,~B., Dettbarn,~C.: \aap~314, 438 (1996) 
\bibitem[\protect\citeauthoryear{Worthey et al.} {1996}]{woa96}  
Worthey,~G., Dorman,~B., Jones,~L.A.: \aj~112, 948 (1996) 
\bibitem[\protect\citeauthoryear{Wyse and Gilmore}{1992}]{wyg92}  
Wyse,~R.F.G., Gilmore,~G.: \aj~104, 144 (1992)
\bibitem[\protect\citeauthoryear{Wyse and Gilmore}{1995}]{wyg95}  
Wyse, R.F.G., Gilmore, G.: \aj~110, 2771 (1995)
\bibitem[\protect\citeauthoryear{Zoccali et al.} {2008}]{zoa08}  
Zoccali,~M., Hill,~V., Lecureur,~A., et al.: \aap~486, 177 (2008) 
\bibitem[\protect\citeauthoryear{Zurek et al.}{1988}]{zua88}  
Zurek,~W.H., Quinn,~P.J., Salmon,~J.K.: \apj~330, 519 (1988) 
\end{thebibliography}
\end{document}